\documentclass[12pt,twoside]{article}
\usepackage{amssymb}
\usepackage{amsmath}
\usepackage{mathabx}
\usepackage{latexsym}
\usepackage{longtable}
\usepackage{epsfig}
\usepackage{physics}
\usepackage{comment}
\usepackage{cancel}
\usepackage{graphicx,bbm,psfrag}
\usepackage{cite}
\graphicspath{{images/}}
\usepackage[usenames,dvipsnames]{color}
\usepackage{appendix}

\newenvironment{proof}{{\bf Proof:}}{\hspace*{\fill}$\Box$}

\newcommand{\beq}{\begin{equation}}
\newcommand{\eeq}{\end{equation}}

\DeclareMathAlphabet{\mathsfit}{T1}{\sfdefault}{\mddefault}{\sldefault}
\SetMathAlphabet{\mathsfit}{bold}{T1}{\sfdefault}{\bfdefault}{\sldefault}

\begin{document}
\begin{titlepage}
\vspace*{5cm}
\begin{center}
\Large{\textbf{Particle Scattering and Fusion for the \smallskip\\Ablowitz-Ladik Chain}}
\bigskip\bigskip\bigskip
\end{center}
\begin{center} 
{\large{Alberto Brollo$^{1,2}$ and Herbert Spohn$^{1,2}$}}\bigskip\bigskip\\
$^1$ Technical University of Munich, CIT, Department of Mathematics,
Boltzmannstraße 3, 85748 Garching, Germany,\smallskip\\
$^2$ Technical University of Munich, Department of Physics, James-Franck-Straße 1, 85748 Garching, Germany.\bigskip\\
E-mail: alberto.brollo@tum.de, spohn@ma.tum.de

\end{center}
\vspace{1cm}
\begin{flushright}
June 24, 2024
\end{flushright}
\vspace{3cm}
\textbf{Abstract}: The Ablowitz-Ladik chain is an integrable discretized version of the nonlinear Schr\"{o}dinger equation. We report on a novel underlying Hamiltonian particle system  
with properties similar to the ones known for the classical Toda chain and Calogero fluid with $1/\sinh^2$
pair interaction. Boundary conditions are imposed such that, both in the distant past and future, particles have a constant velocity. We establish the many-particle scattering for the Ablowitz-Ladik chain and obtain properties known for generic integrable many-body systems. For a specific choice of the chain, real initial data remain real in the course of time. Then, asymptotically, particles   
move in pairs with a velocity-dependent size and scattering shifts are governed by the fusion rule.

\tableofcontents

\end{titlepage}
\section{Introduction}
\setcounter{equation}{0}
\label{sec1}
The hydrodynamic description of many-particle systems is based on the notion of local thermal equilibrium.
When considering integrable many-body systems, by necessity fine-tuned, it is less immediate 
whether such a hydrodynamic description is still meaningful and, if so, what is the form of the hydrodynamic equations. In the context of quantum quenches it was realized that thermal equilibrium should be replaced by generalized Gibbs ensembles (GGE)\cite{PSSV11}. The 2016 breakthrough \cite{CDY16,BCDF16} discovered how to compute the GGE averaged currents. Thereby the equations of generalized hydrodynamics (GHD) could be put forward with confidence.
Since then much progress has been accumulated, as evidenced by lecture notes \cite{D19a,S23}, review articles \cite{GHDbose,DGMSV23}, and special journal volumes \cite{JSTAT24,JPhysA24}, covering classical and quantum many-body systems. Surprisingly, the structure of  GHD 
was unravelled much earlier in the context of soliton gases \cite{EK05,E21,BDE22}.

As understood already very early on, for integrable many-body systems and nonlinear wave equations 
the two-particle scattering shift is a central property. The guiding example is a system of hard rods in one dimension, for which it is evident that $N$-particle scattering shifts are described by a sum of 
two-particle scattering shifts.  For other integrable systems, as for example Toda lattice \cite{M75} and KdV equation \cite{T75,T76}, 
to establish properties of $N$-body scattering  relies on more subtle reasoning. In strong contrast to the hard rod dynamics, the two-particle scattering shift depends now on the incoming velocities. As hallmark of integrability, the asymptotic $N$-particle scattering shifts are still a sum of two-particle scattering shifts\cite{D08}.

When investigating GGEs, it was discovered that the two-particle scattering shift also appears in the variational formula for the generalized free energy. At first sight this may look surprising. But for integrable quantum many-body systems, the Bethe ansatz for the eigenvalues involves already the two-particle phase shift. Thus this shift is expected to reappear in the analysis of the partition function together with the associated free energy functional\cite{F17}. In applications, mostly used are 
the Euler-Lagrange equations for the minimizer of the free energy functional, which in this context are called TBA equations. The TBA equations are written in terms of a linear dressing operator, whose kernel is the two-particle scattering shift. The same structure holds for the classical Toda chain 
\cite{D19,S19} and the Calogero fluid \cite{S23}. But the derivation turns out to be much more indirect.

In our contribution, we consider the defocusing Ablowitz-Ladik (AL) model, which is an integrable discretization of the nonlinear Schr\"{o}dinger equation \cite{AL75,AL76}, see also the book \cite{APT04}. The corresponding discrete wave field, denoted by
$\alpha_1,\dots,\alpha_N$, is complex-valued with the constraint $|\alpha_j| \leq 1$.
Fairly recently, the AL generalized free energy 
has been computed explicitly \cite{GM21,S21}. The corresponding TBA equations involve the dressing operator, $T$, with integral kernel
\begin{equation} 
\label{1.1}
T(\theta,\theta') = 2\log|\mathrm{e}^{\mathrm{i}\theta}-\mathrm{e}^{\mathrm{i}\theta'}|,
\end{equation}
where $\theta,\theta' \in [0,2\pi]$. According to the general scheme, this should be  a two-particle scattering shift. But no Bethe ansatz is at hand and, because of the constraint $|\alpha_j| \leq 1$, there seems to be no obvious notion of scattering. We resolve this intriguing puzzle
by first identifying the underlying particle system governed by canonical evolution equations for position-like and momentum-like variables. Using an exact formula for the time evolution of the AL chain \cite{KN07}, we then establish that the so identified   
$N$-particle trajectories possess the scattering structure of integrable systems.

The AL chain has an infinite tower of conserved charges, and among these an energy-like and a momentum-like conserved charge. The energy-like charge governs the dynamics of the AL chain. But one can also use  the momentum-like charge as  generating the dynamics, which is known as Schur flow. Under the Schur flow, real initial data stay real. Hence the natural phase space is $[-1,1]^N$. Our construction of the underlying $N$-particle trajectories also applies to this case with a constraint on the admissible initial conditions. Highly unexpected, asymptotically in distant past and future, the particles now form bound pairs with a velocity-dependent width. 
The scattering 
between two such pairs follows a fusion rule \cite{T16}, which accounts for the dressing operator found 
earlier in the corresponding TBA equations.

In the following  section, we explain the particle structure. Once having discovered the correct ansatz, the Hamiltonian properties are easily confirmed.  On the level of the Lax matrix, the AL chain carries considerable similarity with the Toda lattice and, 
as a preparation, we briefly recall how Moser \cite{M75} established the $N$-particle scattering for this lattice. In Section  \ref{sec3} we take up the AL chain with the standard Hamiltonian imposing open boundary conditions. In this case the eigenvalues of the Lax matrix are almost surely non-degenerate and so are the rapidities. The Schur flow will be discussed in Section  \ref{sec4}. In this case, while the spectrum of the Lax matrix is still non-degenerate, the rapidities are two-fold degenerate. Some of the more technical material is moved to an Appendix.

\section{Particle dynamics associated with the AL chain}
\setcounter{equation}{0}
\label{sec2a}

A famous integrable wave equation is the nonlinear Schr\"{o}dinger equation (NLS)
in one spatial dimension \cite{APT04}. Its complex-valued wave field, $\psi(x,t)$, $x \in \mathbb{R}$, is governed
by 
\begin{equation}\label{2a.1} 
\mathrm{i} \partial_t\psi  = - \partial_x^2 \psi \pm 2|\psi|^2\psi.
\end{equation}
The $+$ sign is referred  to as defocusing and the $-$ sign as focusing. In our contribution we study only the defocusing version. In the obvious lattice discretization one would consider the complex lattice field $\psi_j(t)$, $j \in\mathbb{Z}$, governed by 
\begin{equation}\label{2a.1a}
\mathrm{i}\frac{d}{dt}\psi_j  = - \psi_{j-1}  + 2 \psi_j - \psi_{j+1}   +2|\psi_j|^2 \psi_{j}.
\end{equation}
But this model is no longer integrable. As discovered by Ablowitz and Ladik \cite{AL75,AL76}, integrability is preserved through the minor modification    
   \begin{equation}\label{2a.1b}
\mathrm{i}\frac{d}{dt}\psi_j  = - \psi_{j-1}  + 2 \psi_j - \psi_{j+1} +  |\psi_j|^2 (\psi_{j-1} + \psi_{j+1}),
\end{equation}
 hence
\begin{equation} \label{2a.1c} 
\mathrm{i}\frac{d}{dt}\psi_j  = - (1 - |\psi_j|^2)(\psi_{j-1} + \psi_{j+1}) + 2 \psi_j.
\end{equation}
Setting $\alpha_j(t) = \mathrm{e}^{2\mathrm{i}t}\psi_j(t)$, one arrives at the standard version
\begin{equation}\label{2a.2} 
\frac{d}{dt}\alpha_j  = \mathrm{i} \rho_j^2 (\alpha_{j-1} + \alpha_{j+1}),\quad \rho_j^2 = 1 - |\alpha_j|^2.
\end{equation}
A natural choice for the  phase space is $\alpha_j \in \mathbb{D}$, where 
$\mathbb{D} = \{z||z| \leq 1\}$ is the complex unit disk. Whenever $\alpha_j (t)$
hits the boundary of $\mathbb{D}$, it freezes and thereby decouples the system. However, a conservation law ensures that, if initially away from the boundary, the solution will stay so forever.
The equations of motion can be derived from the Hamiltonian
\begin{equation}\label{2a.2a} 
H_\mathrm{AL}= -\sum_j \big(\alpha_{j-1} \bar{\alpha}_{j} + \bar{\alpha}_{j-1} \alpha_{j}\big),
\end{equation}
where $\alpha_{j},\bar{\alpha}_{j}$ are viewed as canonically conjugate variables according to the weighted Poisson bracket, 
\begin{equation}\label{2a.2b} 
\{f,g\}_\mathrm{AL} =\mathrm{i} \sum_{j=1}^{N-1} \rho_j^2\big(\partial_{\bar{\alpha}_j} f 
\partial_{\alpha_j} g  - \partial_{\alpha_j} f 
\partial_{\bar{\alpha}_j} g \big).
\end{equation}

To find out about particle trajectories we propose the educated guess 
\beq
\label{2a.3}
\alpha_j=\sqrt{1-\mathrm{e}^{-r_j}}\mathrm{e}^{\mathrm{i}\varphi_j},
\eeq
where $r_j \geq 0$ and $0 \leq \varphi_j < 2\pi$. Simply rewriting 
\eqref{2a.2} in terms of the new variables, $r_j,\varphi_j$, the equations of motion become 
\begin{eqnarray}
\label{2a.4}
&&\hspace{-20pt}\frac{d}{dt}r_j = -2\sqrt{(1-\mathrm{e}^{-r_{j-1}})(1-\mathrm{e}^{-r_j})}\sin(\varphi_{j-1}-\varphi_j) + 2\sqrt{(1-\mathrm{e}^{-r_{j}})(1-\mathrm{e}^{-r_{j+1}})}\sin(\varphi_{j}-\varphi_{j+1}), \nonumber\\
  &&\hspace{-20pt}  \frac{d}{dt}\varphi_j = \frac{\mathrm{e}^{-r_j}}{(1-\mathrm{e}^{-r_j})}\\
  &&\hspace{-20pt}\hspace{20pt}\times\big[\sqrt{(1-\mathrm{e}^{-r_{j-1}})(1-\mathrm{e}^{-r_j})}\cos{(\varphi_{j-1}-\varphi_j)} + \sqrt{(1-\mathrm{e}^{-r_{j}})(1-\mathrm{e}^{-r_{j+1}})}\cos{(\varphi_{j}-\varphi_{j+1})}\big].\nonumber
\end{eqnarray}
 We regard $r_j,\varphi_j$ as canonical coordinates with standard Poisson bracket
\beq
\label{2a.4a}
\{f,g\} = \sum_{j}\big(\partial_{r_j}f\partial_{\varphi_j}g - \partial_{r_j}g\partial_{\varphi_j}f\big).
\eeq
In particularly, $\{r_k,\varphi_j\} = \delta_{kj}$. Expressing $H_\mathrm{AL}$ in the new coordinates, one obtains
\beq
\label{2a.5}
H_\mathrm{p} = -\sum_{j}2\sqrt{(1-\mathrm{e}^{-r_{j-1}})(1-\mathrm{e}^{-r_j})}\cos{(\varphi_{j-1}-\varphi_j)}.
\eeq
Then, indeed, the equations of motion \eqref{2a.4} take the canonical form
\beq
\label{2a.6}
\frac{d}{dt}r_j = \frac{\partial H_\mathrm{p}}{\partial \varphi_j}, \qquad
    \frac{d}{dt}\varphi_j = -\frac{\partial H_\mathrm{p}}{\partial r_j}.
\eeq

So far we did not specify the boundary conditions. For the \textit{closed} chain one would use the obvious periodic boundary conditions. As an integrable system,  since phase space is bounded, it then foliates into invariant tori. Our interest here is 
many-particle scattering, for which purpose particles will travel to infinity, also referred to as \textit{open} chain. For particles  
with labels $ j = 1,\dots, N-1$ the appropriate boundary conditions are obtained by
moving $r_0$ and $r_{N}$ to $+\infty$. The respective phases remain as boundary parameters. Thereby one arrives at
\beq
\label{2a.7}
H_\mathrm{oc} = -\sum_{j=1}^{N}2\sqrt{(1-\mathrm{e}^{-r_{j-1}})(1-\mathrm{e}^{-r_j})}\cos{(\varphi_{j-1}-\varphi_j)},
\eeq
where $\varphi_0,\varphi_N$ are fixed and $r_0 = +\infty$, $r_N = +\infty$.
With these conventions, Eqs. \eqref{2a.4} hold for the indices $ j = 1,\dots, N-1$. 

Up to now, an interpretation in terms of particle positions remained elusive. But with guidance from the Toda lattice, see Section \ref{sec2}, $r_j$ is regarded as positional difference,
\beq
\label{2a.7a}
r_j =q_{j+1} - q_j,
\eeq
with particle positions 
$q_1,\dots,q_{N}$. In the equation of motion for $r_j$ \eqref{2a.4}, one notes that the right hand side is a difference,
which implies
\beq
\label{2a.8}
\begin{split}
&\frac{d}{dt}q_1 = 2\sqrt{1-\mathrm{e}^{-r_1}}\sin(\varphi_0- \varphi_1),\\
&\frac{d}{dt}q_j =  2\sqrt{(1-\mathrm{e}^{-r_{j-1}})(1-\mathrm{e}^{-r_j})}\sin(\varphi_{j-1}-\varphi_j),\quad j = 2,\dots, N-1,\\
&\frac{d}{dt}q_N =  2\sqrt{1-\mathrm{e}^{-r_{N-1}}}\sin(\varphi_{N-1}-\varphi_{N})
\end{split}
\eeq
for fixed $\varphi_0, \varphi_N$. Positions are ordered as $q_N(t) \geq \cdots \geq q_1(t) $,
since $r_j(t) \geq 0$.
One observes that for the total velocity,
\beq
\label{2a.9}
\begin{split}
&P_\mathrm{p} = \frac{d}{dt}\sum_{j=1}^N q_j =  2\Big(\sqrt{1-\mathrm{e}^{-r_1}}\sin(\varphi_0- \varphi_1) + \sum_{j=2}^{N-1}
\sqrt{(1-\mathrm{e}^{-r_{j-1}})(1-\mathrm{e}^{-r_j})}\sin(\varphi_{j-1}-\varphi_j)\\
&\hspace{60pt}+ \sqrt{1-\mathrm{e}^{-r_{N-1}}}\sin(\varphi_{N-1}-\varphi_{N})\Big).
\end{split}
\eeq
$P_\mathrm{p}$ is conserved and hence the center of mass moves inertially. 

Ignoring momentarily boundary conditions, the total velocity can be written as 
\beq
\label{2a.9aa}
P_\mathrm{p} = \sum_j \pi_j
\eeq
with $\pi_j$ the physical velocity of each particle, which  according to \eqref{2a.8} is given by
\beq
\label{2a.9a}
\pi_j = 2\sqrt{(1-\mathrm{e}^{-r_{j-1}})(1-\mathrm{e}^{-r_j})}\sin(\varphi_{j-1}-\varphi_j).
\eeq
On purpose we refer to $\pi_j$ as velocity density and  not as momentum density. Thereby we avoid to construct the Hamiltonian momentum conjugated to $q_j$. Even though $\{r_j,\phi_j\}$ are Hamiltonian conjugated variables, starting from the definition \eqref{2a.7a} of $q_j$, it is not so obvious how to find their conjugated variables. 

Correspondingly the total energy $H_\mathrm{p}$ in \eqref{2a.5} has the energy density
\beq
\varepsilon_j = -2\sqrt{(1-\mathrm{e}^{-r_{j-1}})(1-\mathrm{e}^{-r_j})}\cos(\varphi_{j-1}-\varphi_j), 
\eeq
which then implies the dispersion relation of the system as 
\beq
\label{2a.9b}
\varepsilon_j = \pm\sqrt{1-\pi_j^2}.
\eeq
Note that the map between the angles $\{\varphi_j\}$ and the local momenta $\{\pi_j\}$ is not a bijection
and has to be understood piecewise.

Let us switch back to the $\alpha_j$ variables. The total velocity $P_p$ can be expressed as
\begin{equation}\label{eq:momentum}
P_\mathrm{AL}= -\mathrm{i}\sum_j \big(\alpha_{j-1} \bar{\alpha}_{j} - \bar{\alpha}_{j-1} \alpha_{j}\big).
\end{equation}
By analogy we refer to $P_\mathrm{AL}$ as total momentum.
Instead of $H_\mathrm{AL}$, one can also choose the total momentum  $P_\mathrm{AL}$ as generator of the dynamics. Equipped with the Poisson brackets \eqref{2a.2b}, one arrives at the equations of motion
\beq\label{2a.11}
\frac{d}{dt}\alpha_j  = \rho_j^2 (\alpha_{j+1} - \alpha_{j-1}),\quad \rho_j^2 = 1 - |\alpha_j|^2,
\eeq
which is known as Schur flow. Trivially such dynamics is also integrable.  The equations of motion for the open chain in the $\{r_j,\varphi_j\}$ variables are 
\begin{eqnarray}
\label{2a.10}
    &&\hspace{-20pt}\frac{d}{dt}r_j = -2\sqrt{(1-\mathrm{e}^{-r_{j-1}})(1-\mathrm{e}^{-r_j})}\cos(\varphi_{j-1}-\varphi_j) + 2\sqrt{(1-\mathrm{e}^{-r_{j}})(1-\mathrm{e}^{-r_{j+1}})}\cos(\varphi_{j}-\varphi_{j+1}), \nonumber\\
    &&\hspace{-20pt}\frac{d}{dt}\varphi_j = \frac{-\mathrm{e}^{-r_j}}{(1-\mathrm{e}^{-r_j})}\\
    &&\hspace{4pt}\times\big[+\sqrt{(1-\mathrm{e}^{-r_{j-1}})(1-\mathrm{e}^{-r_j})}\sin{(\varphi_{j-1}-\varphi_j)} + \sqrt{(1-\mathrm{e}^{-r_{j}})(1-\mathrm{e}^{-r_{j+1}})}\sin{(\varphi_{j}-\varphi_{j+1})}\big],\nonumber
\end{eqnarray}
$j =1,\dots, N-1$, with boundary conditions $r_0 = +\infty$, $r_N = +\infty$.  Eq. \eqref{2a.6} remains valid upon substituting $H_\mathrm{p}$ by $P_\mathrm{p}$.

In fact, Schur and AL flow are related through a simple transformation. We consider the AL flow with general initial data $r_j(0),\varphi_j(0)$. Define $\tilde{r}_j(0) = r_j(0)$ and $ \tilde{\varphi}_j(0) = \varphi_j(0) + j\pi/2$. By inserting in \eqref{2a.4}, one confirms that $\tilde{r}_j(t) = r_j(t) $, $\tilde{\varphi}_j(t) = \varphi_j(t) + j\pi/2$  is solution of the Schur flow.

The Schur flow possesses a simplification. If, for all $j$,  $\varphi_j(0) = 0, \pi$, then $\varphi_j(t) = 0,\pi$ for all $t$\footnote{In general, due to the global $U(1)$ symmetry of the system, any straight line passing through the center of the unit disk is an invariant subspace.}. Thereby the dimension of 
phase space is cut in half, which is a set of measure zero in terms of the original phase space.
Very different dynamical behavior might be anticipated. With such constraint, the dynamical variables are $\alpha_j \in [-1,1]$ and they satisfy \eqref{2a.11} with $\alpha_0 = -1$, $\alpha_N = -1$. The Hamiltonian structure is lost. But particle positions can be introduced in analogy to Eqs. \eqref{2a.7a}, \eqref{2a.8}. Of course, by the mapping discussed in the previous paragraph, also the AL flow has a corresponding $N$-dimensional invariant subspace.

The equations of motion \eqref{2a.4} and \eqref{2a.8} allow us to solve the two-body problem and extract the scattering shift as done in Appendix A.
Nevertheless, there seems to be no way to establish on this basis such a fine asymptotic property as $N$-body scattering. To make progress we will return to the initial version \eqref{2a.2}. As to be explained, for this system powerful tools are available. As an introduction to the method, we first recall the case of the Toda lattice which serves as a guiding example.
\section{The open Toda chain}
\setcounter{equation}{0}
\label{sec2}
 To set the scene we start from a generic chain with a repulsive potential
$V_{\mathrm{mec}}$, which is defined on the open interval $]d_0,+\infty[$, $d_0 = -\infty$ included, repulsive, i.e. $V'_{\mathrm{mec}} <0$, diverging to $+\infty$ at $d_0$, and decaying sufficiently rapidly to zero at infinity. We consider $N$ particles moving on the real line, positions and momenta being denoted by $q_j,p_j$, $j = 1,\dots,N$,  $q = (q_1,\dots,q_N) $ and  $p = (p_1,\dots,p_N) $. The dynamics is governed by Newton's equations of motion obtained from the Hamiltonian
 \begin{equation}\label{2.1}
H_{\mathrm{mec}} = \sum_{j = 1}^N\tfrac{1}{2}p_j^2 + \sum_{j=1}^{N-1} V_\mathrm{mec}(q_{j+1} - q_j).
\end{equation}
Initially particles start at $(q,p)$ and evolve in time to $(q(t),p(t))$.
Then, in the distant past and future, particles move freely. On abstract grounds \cite{H89}, one can prove that asymptotic momenta exist, 
 \begin{equation}\label{2.2}
\lim_{t \to \pm\infty} p_j(t) = p_j^\pm,
\end{equation}
and also the, in time,  forward and backwards scattering shifts
 \begin{equation}\label{2.3}
\lim_{t \to \pm\infty} q_j(t)- p_j^{\pm }t = \phi_j^{\pm}, \qquad \phi_j^{\pm} \in \mathbb{R}.
\end{equation}

For large times, particles move along a straight line and hence are asymptotically free.
The scattering data  depend on the  initial conditions. Since the potential is repulsive, if the incoming momenta are ordered as $p_1^- > \dots > p_N^-$, then the outgoing momenta appear in reversed order  $p_1^+ < \cdots < p_N^+$. In fact,  the scattering map is canonical and thus  satisfies  the Poisson bracket relations
  \begin{equation}\label{2.4}
\{p_i^{+},p_j^{+}\} = 0,\quad \{p_i^{+},\phi_j^{+}\} = \delta_{ij},\quad      \{\phi_i^{+},\phi_j^{+}\} =0,
\end{equation}
correspondingly for the past. In scattering coordinates the Hamiltonian trivializes to the one of free particles as 
\begin{equation}\label{2.5}
\sum_{j=1}^N \tfrac{1}{2} (p_j^{\pm})^2.
\end{equation}
The set $\{p_j^+,j=1,\dots,N\}$ are $N$ integrals of motion in involution.  However, they have an intricate dependence on initial data and most likely do not have a density  local in the particle label. Only integrability ensures quasi-local densities of the conserved charges.
For chains with a Hamiltonian of the form \eqref{2.1}, the choice reduces to the fine-tuned potential $V_\mathrm{mec}(x) = \mathrm{e}^{-x}$, which is the well-known Toda chain.

Scattering will be studied for the open Toda chain governed by the Hamiltonian 
\begin{equation}\label{2.7}
 H_\mathrm{to} = \sum_{j=1}^N \tfrac{1}{2}p_j^2 +  \sum_{j=1}^{N-1}a_j^2,
\end{equation}
where we introduced the Flaschka variables 
\begin{equation}\label{2.6}
 a_j  = \mathrm{e}^{-(q_{j+1} - q_j)/2}
\end{equation} 
\cite{M75}. In these coordinates the equations of motion become
\begin{eqnarray}\label{2.8}
&&\hspace{0pt}\frac{d}{dt} a_j = \tfrac{1}{2}a_j(p_j - p_{j+1}),\quad j = 1,\dots,N-1,\nonumber\\[1ex]
&&\hspace{0pt} \frac{d}{dt} p_j = a_{j-1}^2 - a_{j}^2, \quad j = 1,\dots,N, 
\end{eqnarray}
with the boundary conditions $a_0 = 0$, $a_N = 0$. In addition,  we introduce the Lax matrix, $L$, which is the $N\times N$ symmetric Jacobi matrix with $L_{j,j} = p_j$, $L_{j,j+1} = L_{j+1,j} = a_j$, and $L_{i,j}=0$ otherwise. The eigenvalues of $L$ are conserved in time and hence  the asymptotic momenta coincide with the eigenvalues of $L$. 

For incoming and outgoing configurations, particles are labelled by increasing index from left to right and the eigenvalues are ordered as  $\lambda_N>\cdots>  \lambda_1$. This  implies
\begin{equation}\label{2.9}
p_j^{+} = \lambda_j,\qquad  p_j^{-} =  \lambda_{N-j+1}.
\end{equation}
Through scattering the order of asymptotic momenta is merely reversed. Therefore it is convenient to introduce the notion of quasi-particles. The quasi-particle with label $j$ moves with momentum $\lambda_j$ and its trajectory coincides with the one of $q_{N-j+1}$ in the far past and with $q_j$ for the far future. Hence the system evolves in time as an ideal gas, for which particles maintain their momenta, apart for shifts due to collisions with other quasi-particles. Interactions are encoded 
by the structure of the quasi-particle scattering shift.
The natural intrinsic property is then the \textit{relative scattering shift} of the $j$-th quasi-particle defined by 
\begin{equation}\label{2.10}
\Delta_j= \phi^+_{j}  - \phi^-_{N-j+1}.
\end{equation}
To be brief,  the extra ``relative" is mostly omitted.
From solving the two-body problem, $N=2$, the Toda two-particle scattering shift  is
obtained as
\begin{equation}\label{2.11}
2\log|p_2 - p_1|,
\end{equation}
in case of incoming momenta with $p_1 > p_2$. 

Scattering of $N$ particles requires more sophisticated tools and,
for later purposes, it is useful to recall the basic strategy \cite{M75}. 
From \eqref{2.2}, \eqref{2.3} it follows that 
\begin{equation}\label{2.13} 
\sum_{j=1}^N \Delta_j = 0.
\end{equation}
Thus it suffices to study the limit related to $\Delta_{j+1} - \Delta_j$, $j = 1,\dots,N-1$. According to \eqref{2.3}, \eqref{2.9} the Flaschka variables have the asymptotics
\begin{equation}\label{2.14} 
a_j(t) = \xi_j\mathrm{e}^{-(\lambda_{j+1}-\lambda_{j})t/2},\hspace{5pt} t  \to +\infty,\qquad
a_j(t) = \zeta_j\mathrm{e}^{(\lambda_{N-j+1}-\lambda_{N- j})t/2},\hspace{5pt} t  \to -\infty,
\end{equation}
with coefficients
 \begin{equation}\label{2.15}
 \xi_j = \exp((\phi_j^+ - \phi_{j+1}^+)/2),\qquad  \zeta_j= \exp((\phi_j^- - \phi_{j+1}^-)/2).
\end{equation}
Thus it is natural to investigate the limit
\begin{equation}\label{2.16} 
\lim_{t \to +\infty} a_{j}(t)a_{N-j}(-t)\mathrm{e}^{(\lambda_{j}-\lambda_{j+1})t} = \xi_{j}\zeta_{N-j} = \mathrm{e}^{(\Delta_{j} - \Delta_{j+1})/2}.
\end{equation}
As proved in \cite{M75} the limit equals
 \begin{equation}\label{2.17} 
\xi_{j}\zeta_{N-j}=\frac{\prod_{1\leq i< j}(\lambda_j-\lambda_i)}{\prod_{j < i\leq N}(\lambda_i-\lambda_j)}\cdot
\frac{\prod_{ j+1 < i \leq N}(\lambda_i-\lambda_{j+1})}{\prod_{1 \leq i< j+1}(\lambda_{j+1} -\lambda_i)},
\end{equation}
where empty products are set equal to $1$. Squaring and
taking the logarithm on both sides yields
\begin{eqnarray}\label{2.18} 
&&\hspace{-60pt}\Delta_{j} - \Delta_{j+1} = \log|\xi_{N- j} \zeta_{j}|^2 
 = \sum_{1\leq i< j} 2\log|\lambda_i - \lambda_j| -  \sum_{j < i\leq N}2\log|\lambda_i -\lambda_j|  \nonumber\\
&&\hspace{82pt}
 - \sum_{1 \leq i< j+1} 2\log|\lambda_i - \lambda_{j+1}| + \sum_{ j+1 < i \leq N}2\log|\lambda_i - \lambda_{j+1}|.
\end{eqnarray}
It is convenient to introduce the sign convention
\beq\label{2.19}
\phi_{\mathrm{to}}(\lambda_j,\lambda_k)=
\begin{cases}
    2\log{\abs{\lambda_j-\lambda_k}} \qq{for} \lambda_j<\lambda_k,    \\
    -2\log{\abs{\lambda_j-\lambda_k}} \qq{for} \lambda_j>\lambda_k.    \\
\end{cases}
\eeq
Then
the $N$-particle shift $\Delta_j$ is simply the sum of  two-particle scattering shifts as
\begin{equation}\label{2.20}
\Delta_j = \sum_{1 \leq k  \leq N,k\neq j} \phi_{\mathrm{to}}(\lambda_j,\lambda_k).
\end{equation}
While the actual interactions are much more intricate, as regards to scattering one can view this asymptotics as resulting from
a succession of two-body collisions just as for a system of hard rods, keeping
in mind that the scattering shift depends on the incoming velocities. 

The Toda lattice can be viewed as a discretized wave equation. In the context of integrable nonlinear 
continuum wave equations in one dimension, a comparable scattering theory has been established
earlier. Historically, in 1834,
two-body scattering was first observed for water waves in shallow channels and subsequently well modelled by the Korteweg-de Vries (KdV) equation in 1895. 
These particular type of solutions became known as solitons. Integrability of the KdV wave equation  started to be  understood in the early 1960ies \cite{K67,T75,T76}. In particular $N$-soliton solutions were constructed whose scattering has the same properties as  explained above for the Toda lattice, of course substituting the appropriate two-soliton shift.  Thus as regards to asymptotic scattering, $N$ Toda particles behave just like $N$ solitons of the KdV equation (and other many-body integrable systems).
 
\section{The open AL chain, non-degenerate rapidities}
\setcounter{equation}{0}
\label{sec3}
\subsection{$N$-particle scattering}
\label{sec3.1}
The defocusing AL chain is governed by 
\begin{equation}\label{3.1} 
\frac{d}{dt}\alpha_j  = \mathrm{i} \rho_j^2 (\alpha_{j-1} + \alpha_{j+1}),\quad \rho_j^2 = 1 - |\alpha_j|^2.
\end{equation}
The natural phase space is $\alpha_j \in \mathbb{D}$ with the complex unit disk $\mathbb{D} = \{z| \abs{z} \leq 1\}$. In principle, whenever a particular $\alpha_j (t)$ hits the boundary of $\mathbb{D}$, it freezes and thereby decouples the system. For the \textit{closed} chain one considers the segment $j=1,\dots,N$ and imposes
the obvious periodic boundary conditions. Then the motion stays away from the boundary of $\mathbb{D}^N$ and the phase space is foliated into invariant tori with quasi-periodic motion. However, for scattering theory one has to study the \textit{open} chain, for which one imposes the boundary conditions $\alpha_0 = -1$ and 
$\alpha_N = \exp(\mathrm{i}\phi_N)$ with a fixed phase $\phi_N \in [0,2 \pi)$. In principle, one could allow a general phase also for $\alpha_0$. Because of the global $U(1)$ symmetry of the model, without loss of generality, the phase of $\alpha_0$ can be fixed to $\pi$. The equations of motion then read  
\begin{equation}\label{3.2} 
\frac{d}{dt}\alpha_j  = \mathrm{i} \rho_j^2 (\alpha_{j-1} + \alpha_{j+1}),\quad j = 1,\dots, N-1,\quad 
\alpha_0 = -1,\quad \alpha_N = \exp(\mathrm{i}\phi_N).
\end{equation}
For simplicity we will require $N$ to be even.

As discovered by Nenciu \cite{N05}, the open AL chain has a Lax matrix, which has a structure known as  Cantero-Moral-Vel\'{a}zquez (CMV) matrix. The basic building blocks are $2\times 2$ matrices,
denoted by
\begin{equation}\label{3.3} 
\Xi_j =
\begin{pmatrix}
\bar{\alpha}_j & \rho_j \\
\rho_j & -\alpha_j  \\
\end{pmatrix}.
\end{equation}
One constructs the $N\times N$ matrices, $L,M$, with
\begin{equation}\label{3.4} 
 L = \mathrm{diag}(\Xi_1,\Xi_3,\dots, \Xi_{N-1}), \quad \big( M\big)_{i,j = 2,\dots,N-1} = \mathrm{diag}(\Xi_2,\Xi_4,\dots, \Xi_{N-2}),
\end{equation}
together with $M_{1,1} =  - \alpha_0$,  $M_{N,N} = \bar{\alpha}_N$, and $M_{i,j} = 0$ otherwise.  The CMV matrix associated to the field configuration $\alpha_1,\dots,\alpha_{N}$ is then given by 
\begin{equation} \label{3.5}
C = L M.
\end{equation}
$C$ is the Lax matrix of the AL chain. Obviously, $L,M$ are unitary and so is $C$. The determinant equals $\det(C) = \alpha_0\bar{\alpha}_N$. The  eigenvalues of $C$ are denoted by $\{z_j, j = 1,\dots,N\}$, $|z_j| = 1$. Of course, $L,M,C$ and the eigenvalues depend on $N$, which is suppressed in our notation. 

The conserved total charges are defined through the Lax matrix as
\begin{equation} \label{3.6}
Q^{[n]} = \mathrm{tr}\big[C^n\big],
\end{equation}
$n= 1,2,\dots$. Clearly, the charges have a strictly local density. They are 
complex-valued, while the physical charges are usually taken as real and imaginary part.
Twice the real part of $Q^{[1]}$ plays the role of the energy, 
\begin{equation} \label{3.7}
H =  \mathrm{tr}[C + C^*] = -\sum_{j=0}^{N-1}(\alpha_j \bar{\alpha}_{j+1} + \bar{\alpha}_j \alpha_{j+1}).
\end{equation}
According to the weighted Poisson brackets \eqref{2a.2b}
\begin{equation}
\{f,g\}_\mathrm{AL} =\mathrm{i} \sum_{j=1}^{N-1} \rho_j^2\big(\partial_{\bar{\alpha}_j} f 
\partial_{\alpha_j} g  - \partial_{\alpha_j} f 
\partial_{\bar{\alpha}_j} g \big),
\end{equation}
one checks that indeed
\begin{equation} \label{3.9}
\frac{d}{dt}\alpha_j = \{\alpha_j,H\}_\mathrm{AL}.
\end{equation}
As pointed out already in \eqref{eq:momentum}, the charge $P =\mathrm{i}\, \mathrm{tr}(C - C^*)$ will play the role of total momentum. In analogy to the Toda lattice, we will show that the correct choice for the rapidities are the eigenvalues of $P$, i.e. $\lambda_j=-2\Im(z_j)$ with $\{z_j\}$ the eigenvalues of $C$.

For the Toda lattice, the eigenvalues of the Lax matrix are known to be non-degenerate and hence can be strictly ordered on the real line. Such a property does not hold for AL.  As discussed in \cite{S23}, Section 12.3, under the measure
\begin{equation} \label{3.9a}
\mathrm{d}\phi_N \prod_{j=1}^{N-1}\mathrm{d}^2\alpha_j(\rho_j^2)^{-1} (\rho_j^2)^{\mathsfit{P}(N-j)},\quad \mathsfit{P}>0,
\end{equation}
there is an explicit formula for the joint distribution of eigenvalues of the Lax matrix,
which has a density relative to $\mathrm{d}\lambda_1\dots \mathrm{d}\lambda_N$.
Hence the set $\{\lambda_i = \lambda_j\}$ has zero measure with respect  to the flat volume measure
\begin{equation} \label{3.9b}
\mathrm{d}\phi_N \prod_{j=1}^{N-1}\mathrm{d}^2\alpha_j.
\end{equation}
Non-degeneracy holds up to a set of volume measure zero. In the following 
we will ignore the exceptional set and strictly order as
 $\lambda_N > \dots >\lambda_1$.
\begin{figure}
    \centering
    \includegraphics[width=0.5\textwidth]{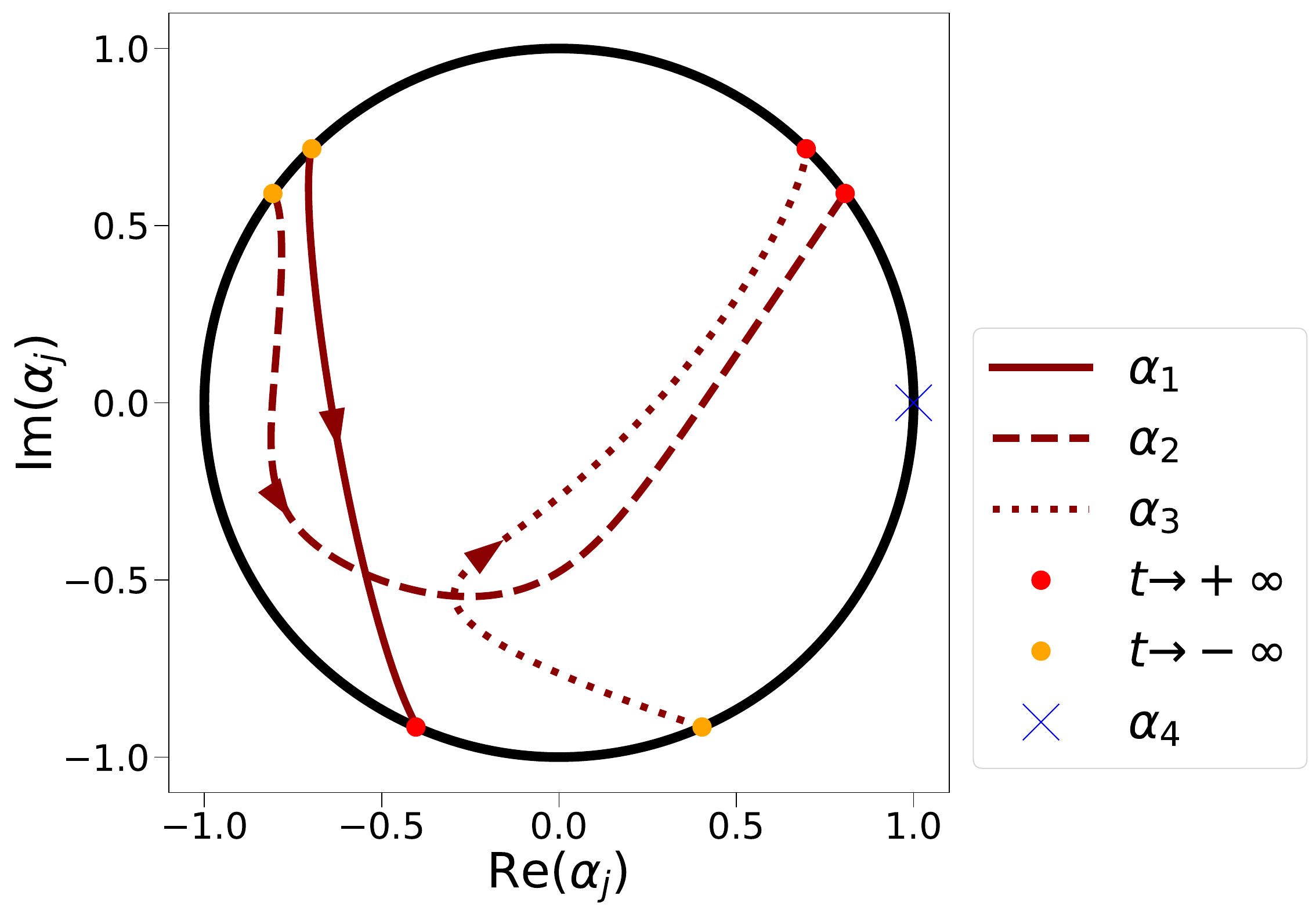}
    \caption{Numerical realization of phase space trajectories for the Ablowitz-Ladik chain with four sites, $N=4$,
    and boundary conditions $\alpha_0=-1$ and $\alpha_4=1$. The black circle is the border of the unit disk $\mathbb{D}$, while the dark red lines are the $\alpha_j(t)$ trajectories except for $\alpha_4(t) = 1$. 
    The initial data are chosen inside the unit disk and the trajectories are computed via \eqref{3.1}. For large $|t|$ they reach the phase space boundary and freeze.
    The dots on the circle are the asymptotic values of the trajectories, orange for the far past and red  for the far future. }
    \label{fig:ALphasespace}
\end{figure}
As proved in \cite{KN07}, see also Appendix B, if the set of rapidities, $\{\lambda_j = -2\Im(z_j), j = 1,\ldots, N\}$, has no multiplicity, then 
\begin{equation} \label{3.10}
\begin{split}
&\lim_{t \to +\infty} \alpha_j(t)=(-1)^{j-1} \Bar{z}_1\dots\Bar{z}_{j}
\qty[1+\xi_{j}\mathrm{e}^{-(\lambda_{j+1}-\lambda_{j})t}+\order{\mathrm{e}^{-\gamma t}}] \\
&\lim_{t \to -\infty} \alpha_j(t) = (-1)^{j-1} \Bar{z}_N\dots\Bar{z}_{N-j+1}
\qty[1+\xi_{j}\mathrm{e}^{(\lambda_{N-j+1}-\lambda_{N-j})t}+\order{\mathrm{e}^{\delta t}}],  \\
\end{split}
\end{equation}
for $j=1,\ldots,N$, where $\gamma>(\lambda_{j+1}-\lambda_{j})>0$ and $\delta>(\lambda_{N-j+1}-\lambda_{N-j})>0$. Notice that, according to the ordering of rapidities, the exponents are strictly negative, hence $|\alpha_j(+\infty)|=|\alpha_j(-\infty)|=1$.
For a numerical simulation of the dynamics inside the phase space check Figure \ref{fig:ALphasespace}.
 The prefactors of  the exponentials are
\begin{equation}\label{3.11}
\begin{split}
&\xi_j=(z_{j}\Bar{z}_{j+1}-1)\frac{\mu_{j+1}}{\mu_{j}}\prod_{k=1}^{j-1}\abs{\frac{z_{j+1}-z_k}{z_{j}-z_k}}^2, \\
&\zeta_j=(z_{N-j+1}\Bar{z}_{N-j}-1)\frac{\mu_{N-j}}{\mu_{N-j+1}}\prod_{k=N-j+2}^{N}\abs{\frac{z_{N-j}-z_k}{z_{N-j+1}-z_k}}^2,
\end{split}
\end{equation}
where empty products are set to 1. The coefficients $\{\mu_1,\ldots,\mu_N\}$ appearing in the above formula are the spectral measures of the Lax matrix $C$ for the vector $e_1=(1,0,\ldots,0)$. More explicitly, if $\Pi_j=\ket{z_j}\bra{z_j}$ denotes the projector to the eigenspace of the $j$-th eigenvalue, one has
\begin{equation}\label{3.12}
    \mu_j=\expval{\Pi_j}{e_1} = \abs{\ip{e_1}{z_j}}^2.
\end{equation}
The Lax matrix, $C(t)$, is time-dependent through the motion of $\alpha_j(t)$'s and so is the spectral measure $\{\mu_j(t)\}$. When this dependence is not specified, we are referring to the one at $t=0$.

In equation \eqref{2a.9a}, the velocities of the underlying particles are defined as $\pi_j(t)=2\Im(\alpha_{j-1}(t)\bar{\alpha_j}(t))$. Eqs. \eqref{3.10} tell us that $\alpha_{j-1}(t)\bar{\alpha_j}(t)\to-z_j$ for $t\to+\infty$. As a result, we have $\pi_j(+\infty)=\lambda_j=-2\Im{z_j}$, corroborating our guess for the rapidities.

Moreover, from \eqref{3.10} one infers that, in contrast to the closed chain,  every $\alpha_j(t)$ reaches the boundary of $\mathbb{D}$ exponentially fast. Thus a natural object for scattering is $\log \rho_j^2(t)$ which  asymptotically is linear in $t$. In a rough sense, $\alpha_j(t)$ should correspond to the Flaschka variable $a_j(t)$ of the Toda model which, as discussed already in Section \ref{sec2a}, amounts to setting
 \begin{equation} \label{3.13}
r_j(t) = -\log \rho_j^2(t). 
\end{equation}
Our task is to establish the long time asymptotics including scattering shifts.

From \eqref{3.10}, one easily finds that for $t\to+\infty$,
\beq\label{3.14}
\begin{split}
&\rho_j^2(t) = -2\Re{\xi_{j}}e^{-(\lambda_{j+1}-\lambda_{j})t} + \order{e^{-\gamma t}}, \\
&\rho_j^2(-t) = -2\Re{\zeta_{j}}e^{-(\lambda_{N-j+1}-\lambda_{N-j})t} + \order{e^{-\delta t}}, \\
\end{split}
\eeq
resulting in the analogue of the Moser limit  \eqref{2.16},
\beq\label{3.15}
\lim_{t\to+\infty} \rho_{j}^2(t)\rho_{N-j}^2(-t)e^{2(\lambda_{j+1}-\lambda_j)t}=4\Re{\xi_{j}}\Re{\zeta_{N-j}}=\exp(\Delta_{j}-\Delta_{j+1}).
\eeq
More explicitly, the limit \eqref{3.15} is given by
\beq\label{3.17}
4\Re{\xi_{j}}\Re{\zeta_{N-j}}=\abs{z_{j+1}-{z}_{j}}^4\prod_{k=1}^{j-1}\abs{\frac{z_{j+1}-z_k}{z_{j}-z_k}}^2
\prod_{k=j+2}^{N}\abs{\frac{z_{j}-z_k}{z_{j+1}-z_k}}^2.
\eeq
Taking the log, one thus obtains
\beq\label{3.18}
\begin{split}
\Delta_{j} -\Delta_{j+1} = &-\sum_{k<j}\log{\abs{z_j-z_k}^2}+\sum_{k>j}\log{\abs{z_j-z_k}^2} \\
&+\sum_{k<j+1}\log{\abs{z_{j+1}-z_k}^2}-\sum_{k>j+1}\log{\abs{z_{j+1}-z_k}^2}.
\end{split}
\eeq

To establish a connection between this limit and the relative scattering shift, one first notes that
\begin{equation}\label{3.16}
\sum_{j=1}^N \Delta_j = 0,
\end{equation}
which is a direct consequence of the linear motion of the center of mass as ensured by the 
conservation of total momentum.
Following the discussion of the Toda lattice, it is convenient to define
\beq\label{3.19}
\phi_{\mathrm{al}}(\lambda_j,\lambda_k)=
\begin{cases}
    -\log{\abs{z_j-z_k}}^2 \qq{for} \lambda_j<\lambda_k,    \\
    \log{\abs{z_j-z_k}}^2 \qq{for} \lambda_j>\lambda_k    \\
\end{cases}
\eeq
and hence
\begin{equation}\label{3.20}
\Delta_j = \sum_{1 \leq k  \leq N,k \neq j} \phi_{\mathrm{al}}(\lambda_j,\lambda_k).
\end{equation}
Comparing with Eq. \eqref{2.20}, the $N$-particle scattering shift can again be written as a sum over two-particle scattering shifts.
\begin{figure}
    \centering
    \includegraphics[width=\textwidth]{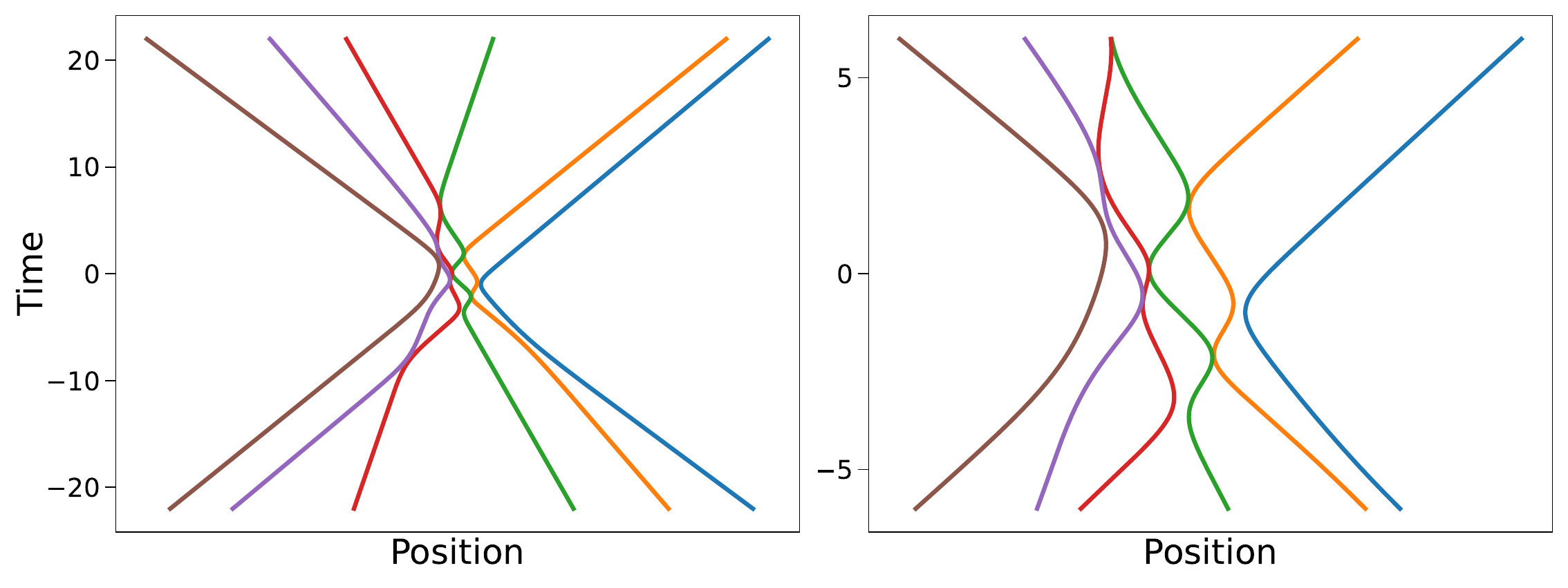}
    \caption{Example of particle trajectories for the Ablowitz-Ladik chain. On the left displayed is large time scale. The quasi-particle picture appears neatly. On the right represented is the short time scale with its intricate particle
    dynamics. Near collisions the motion is parabolic in approximation.}
    \label{fig:ALscattering}
\end{figure}
%
%
\subsection{Numerical results}
We numerically solve the open chain \eqref{3.2} and determine the particle dynamics. From $\alpha_j(t)$, the stretches are obtained through $r_j(t)=-\log(1-\abs{\alpha_j(t)}^2)$. The particle trajectory $q_1(t)$ then follows from \eqref{2a.8}, whereas all other particle trajectories are reconstructed from the $r_j(t)$'s.
In Figure \ref{fig:ALscattering} presented is the motion of six particles, which corresponds to a chain with $N=6$.  In the left panel displayed is a large time scale, at which the quasi-particle picture emerges convincingly. Before and after scattering, particles have the same set of velocities, even though the velocity of a specific particle has changed. Moreover, the scattering shift can be observed, at least qualitatively. A very clear example is the line green in the past and red in the future.
In the right panel, corresponding to a short time scale, observed is the intricate pattern of two-particle collisions.
In contrast to the Toda lattice, particles may touch, but never cross, since $r_j=q_{j+1}-q_j\geq 0$. Inspecting Figure \ref{fig:ALscattering}, particle trajectories at a nearby collision look parabolic. This can be understood from the equations of motion \eqref{2a.4} by expanding near $r_j(t) = 0$. Then
$\dot{r}_j(t)= c(t)\sqrt{r_j(t)}$ with some coefficient $c(t)$ which depends on the neighboring phases and positional differences.
Assuming $c(t)$ to be slowly varying over the time span under consideration, the solution is 
$r_j(t) = \pm(c/2)^2(t+b)^2$. The exact solution of the two-body problem is provided in Appendix A.

\subsection{Comparing with density of states and TBA equation}
\label{sec3.2}
The two-particle scattering shift reappears in the TBA equation.
Actually such connection can be detected already for the density of states of the unitary Lax matrix with matrix elements distributed according to some generalized Gibbs ensemble (GGE). According to equilibrium statistical mechanics, the GGE is first defined in terms of the closed chain, no scattering in sight. Hence one starts from the volume $[1,\ldots,N]$
with periodic boundary conditions. Then the matrix $L$ remains unchanged, while
its shift by one site yields the matrices $M^\diamond$, i.e. the matrix $M$ and $M^\diamond$ differ only by $(M^\diamond)_{1,N} = \rho_N $, $(M^\diamond)_{N,1} = \rho_N $. 
As before, the Lax matrix 
is $C^\diamond = LM^\diamond$. According to the rules, the GGE is constructed from the a priori measure 
weighted with a Boltzmann factor. This factor is the exponential of an arbitrary linear combination of the conserved charges, which can be written compactly as 
$\mathrm{tr}[\mathsfit{V}(C^\diamond)]$. Hence the GGE is given by
\begin{equation} \label{3.26}
\frac{1}{Z} \exp\big(- \mathrm{tr}[\mathsfit{V}(C^\diamond)]\big) (\rho^2_j)^{\mathsfit{P}-1}\prod_{j=1}^{N}\mathrm{d}^2\alpha_j.
\end{equation}
The second factor is the a priori measure for the closed chain. The pressure $\mathsfit{P}$, $\mathsfit{P}>0$, controls the size of $|\alpha_j|$. 
The confining potential, $\mathsfit{V}$, is a real-valued function
of $\mathrm{e}^{\mathrm{i}\theta}$, which should be thought of as a generalized chemical potential. For thermal equilibrium $\mathsfit{V}(\mathrm{e}^{\mathrm{i}\theta})= \beta \cos \theta$ with $\beta$ the inverse temperature. Of interest is  density of states for 
$C^\diamond$ in the limit $N \to \infty$, when viewed as a random matrix under the GGE \eqref{3.26}.

The analysis of GGEs is based on allowing for a minor
modification 
\cite{S23,S21, GM21}. To explain, we return to the open system and impose a linearly varying pressure which leads to the GGE
\begin{equation} \label{3.27}
 \frac{1}{Z} \exp\big(- \mathrm{tr}[\mathsfit{V}(C)]\big) \mathrm{d}\phi\prod_{j=1}^{N-1}(\rho_j^2)^{-1} (\rho_j^2)^{\mathsfit{P}(N-j)/N}  \mathrm{d}^2\alpha_j. 
\end{equation}
The pressure has slope $1/N$. 
For this model the joint distribution of the $N$ eigenvalues  is given by 
\beq \label{3.28}
 \frac{1}{Z}
\exp\Big( - \sum_{j=1}^N \mathsfit{V}(\mathrm{e}^{\mathrm{i}\theta_j}) + \mathsfit{P} \frac{1}{N} \sum_{i,j=1,i\neq j}^N  \log|\mathrm{e}^{\mathrm{i}\theta_i} - \mathrm{e}^{\mathrm{i}\theta_j} |\Big) \mathrm{d}\theta_1\ldots \mathrm{d}\theta_N.
\eeq
In statistical mechanics the probability distribution \eqref{3.28}
is known as circular log-gas\cite{F10}. Since the coupling strength is proportional to $1/N$, it is the mean-field version of the log-gas. We note that the two-particle scattering shift appears as interaction potential for the phases.

With this input the existence of the infinite volume limit of the GGE  has been proved \cite{MM23}. The correlations of limit measure have exponential decay. There is a free energy functional, which has a unique minimizer. Evaluated at this minimizer yields the free energy of \eqref{3.26}.  
The corresponding Euler-Lagrange, or TBA, equations read
\beq\label{3.29}
\epsilon(\theta) = V(\theta) - \mu - 2\int_{0}^{2\pi} \mathrm{d}\theta'\log|\mathrm{e}^{\mathrm{i}\theta}-\mathrm{e}^{\mathrm{i}\theta'}|
\mathrm{e}^{-\varepsilon(\theta')}.
\eeq
Here $V(\theta) = \mathsfit{V}(\mathrm{e}^{\mathrm{i}\theta})$ and $\varepsilon(\theta)$ is the quasi-energy, defined through the Maxwell-Boltzmann parametrization of the number density, $\rho_\mathrm{n}(\theta)=\mathrm{e}^{-\varepsilon(\theta)}$. From the solution of TBA one infers the density of states.
As central input, TBA contains the \textit{dressing  operator} $T$, which is
defined by
\beq\label{3.30}
Tf(\theta) =  
 2\int_{0}^{2\pi} \mathrm{d}\theta'\log|\mathrm{e}^{\mathrm{i}\theta}-\mathrm{e}^{\mathrm{i}\theta'}|f(\theta').
\eeq
Comparing with Eq. \eqref{3.19}, although there is no Bethe ansatz at our disposal, the remarkable conclusion is that the kernel of the dressing operator agrees with the two-particle scattering shift.


\section{The open Schur flow, two-fold degenerate rapidities}
\setcounter{equation}{0}
\label{sec4}
Because of the Hamiltonian structure defined through the weighted Poisson bracket \eqref{2a.2b}, instead of the Hamiltonian $H$ of \eqref{3.7}, any other conserved field will generate an integrable dynamics. In fact also any linear combination will do, which leads to a general class of Hamiltonians of the form
\beq\label{4.0} 
H_\mathsfit{f} = \mathrm{Im}\, \mathrm{tr}[\mathsfit{f}(C)]
\eeq
with $\mathsfit{f}(z)$ some finite polynomial of $z$ with complex coefficients. The AL chain is the case $\mathsfit{f}(z) = 2\mathrm{i}z$. As established in \cite{KN07}, the corresponding rapidities are $\lambda_j=\Re(z_j\mathsfit{f}'(z_j))$ with $z_j$ the eigenvalues of the Lax matrix $C$. For the AL Hamiltonian, $\mathsfit{f}$ is linear and therefore the rapidities are the eigenvalues of the total momentum $P = \mathrm{i}\,\mathrm{tr}(C - C^*)$. Thus a
second natural choice for the dynamics is the Schur flow generated  by $P$, corresponding to $\mathsfit{f}(z) = -2z$.

\subsection{$N$-particle scattering}
\label{sec4.1}
The dynamics is generated by minus the total momentum, 
\beq\label{4.1}
  P =  \mathrm{i}\,\mathrm{tr}[C-C^*],
\eeq
resulting in the equations of motion
\beq\label{4.1a}
\frac{d}{dt}\alpha_j = \rho_j^2(\alpha_{j+1}-\alpha_{j-1}), \quad \rho_j^2 = 1 - \alpha_j^2.
\eeq
Of course, the conserved charges remain as before and so do GGEs, the generalized free energy,
and TBA. Also the $N$-particle scattering theory is not altered, except for the fact that the rapidities are now given by $\lambda_j = -2\mathrm{Re}(z_j)$. In particular, the results of Section 4 can be applied also to this case provided the definition of rapidities is adjusted and their non-degeneracy is maintained.

Such considerations overlook a very peculiar property of Eq. \eqref{4.1a}. 
Obviously, if initial data are real, then they stay real for all times. 
Thus the phase space can be reduced to $[-1,1]^N$. With respect to $\mathbb{D}^N$ this is 
a set of zero measure and the analysis of the previous section has to be redone from scratch.  
As before, the closed chain corresponds to periodic boundary conditions, while 
the open chain is defined by the boundary conditions $\alpha_0=-1$ and $\alpha_N=-1$. The latter could also be $\alpha_N=1$ and will be commented upon at the end of this section.
Thus we now study the real-valued field, $\alpha_j$, governed by 
\beq\label{4.2}
\frac{d}{dt}\alpha_j = \rho_j^2(\alpha_{j+1}-\alpha_{j-1}), \quad j=1,...,N-1, \quad \alpha_0 = -1,\quad \alpha_N = -1,
\eeq
where $\alpha_j \in [-1,1]$. This system is known as open Schur flow \cite{G06}. Since the Schur flow refers to a particular class of initial conditions, it is not immediately clear which properties are inherited from the underlying Hamiltonian structure.
\begin{figure}
    \centering
    \includegraphics[width=\textwidth]{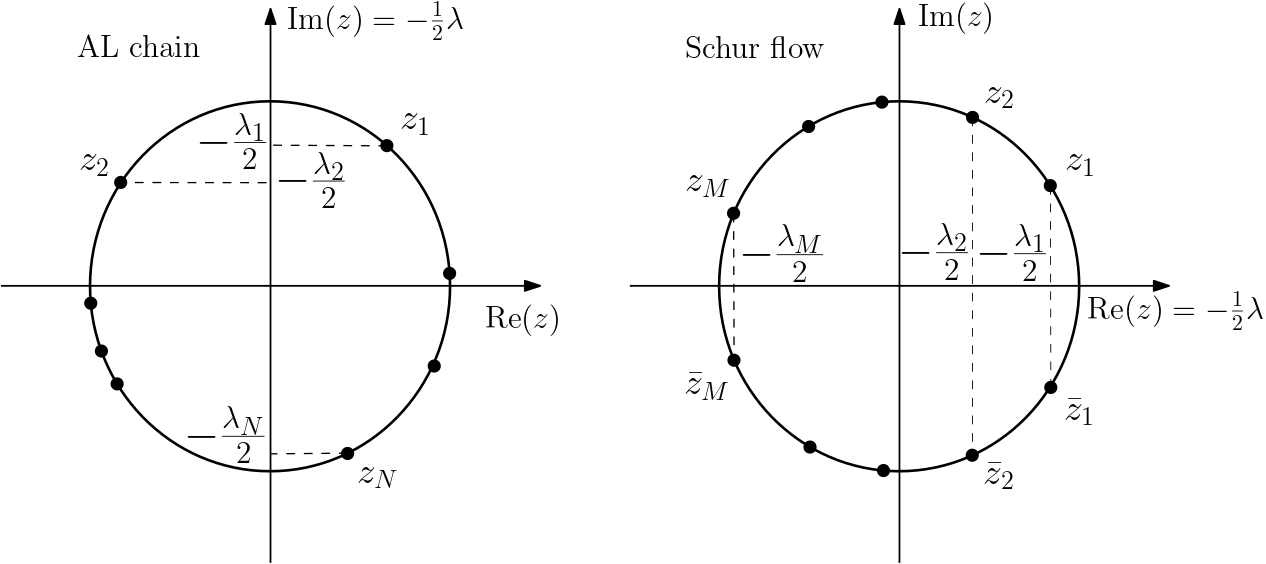}
    \caption{Eigenvalue distribution on the unit circle for the AL chain and Schur flow. On the left displayed is the AL chain. The eigenvalues $z_k=e^{i\theta_k}$, with $\theta_k\in[0,2\pi)$, are labelled according to the rule $\lambda_N>\cdots>\lambda_2>\lambda_1$, where $\lambda_k=-2\sin(\theta_k)$. On the right, displayed are the reflection-symmetric eigenvalues of the Lax matrix for the Schur flow. Each $z_\ell=e^{i\theta_\ell}$, with $\theta_\ell\in[0,\pi]$, has its conjugated $\bar{z}_\ell$ with the same rapidity $\lambda_\ell=-2\cos(\theta_\ell)$. The ordering is $\lambda_M>\cdots>\lambda_2>\lambda_1$.}
    \label{fig:chains}
\end{figure}

The Lax matrix $C$ is unitary, as before, with the further property of being real.  Thus, its eigenvalues are complex conjugated phases $\{\mathrm{e}^{\mathrm{i}\theta_1},\mathrm{e}^{-\mathrm{i}\theta_1},\ldots,\mathrm{e}^{\mathrm{i}\theta_M},\mathrm{e}^{-\mathrm{i}\theta_M}\}$ with $M=N/2$ and $\theta_j\in[0,\pi]$. 
Since  energy and momentum have been swapped, the rapidities become the real part of the eigenvalues, $\lambda_j = -2\cos\theta_j$. The rapidities are two-fold degenerate and strictly ordered as
\beq\label{4.3}
\lambda_M>\cdots> \lambda_1, 
\eeq
except for a set of measure zero with respect to the volume measure $\mathrm{d}\alpha_1\dots
\mathrm{d}\alpha_{N-1}$. For display and comparison with the conventional AL chain, take a look at Figure \ref{fig:chains}.

To determine the two-particle scattering shift required is the asymptotic behaviour. As for the Ablowitz-Ladik chain, we rely on \cite{KN07}. Nevertheless, due to degeneracy, some extra work has to be done. For convenience of the reader, the main theorems of Killip and Nenciu, plus our additional results, are stated in Appendix B. The main novelty comes from the distinction between $\alpha_j$'s labeled by either even or odd index.
\begin{figure}
    \centering
    \includegraphics[width=0.6\textwidth]{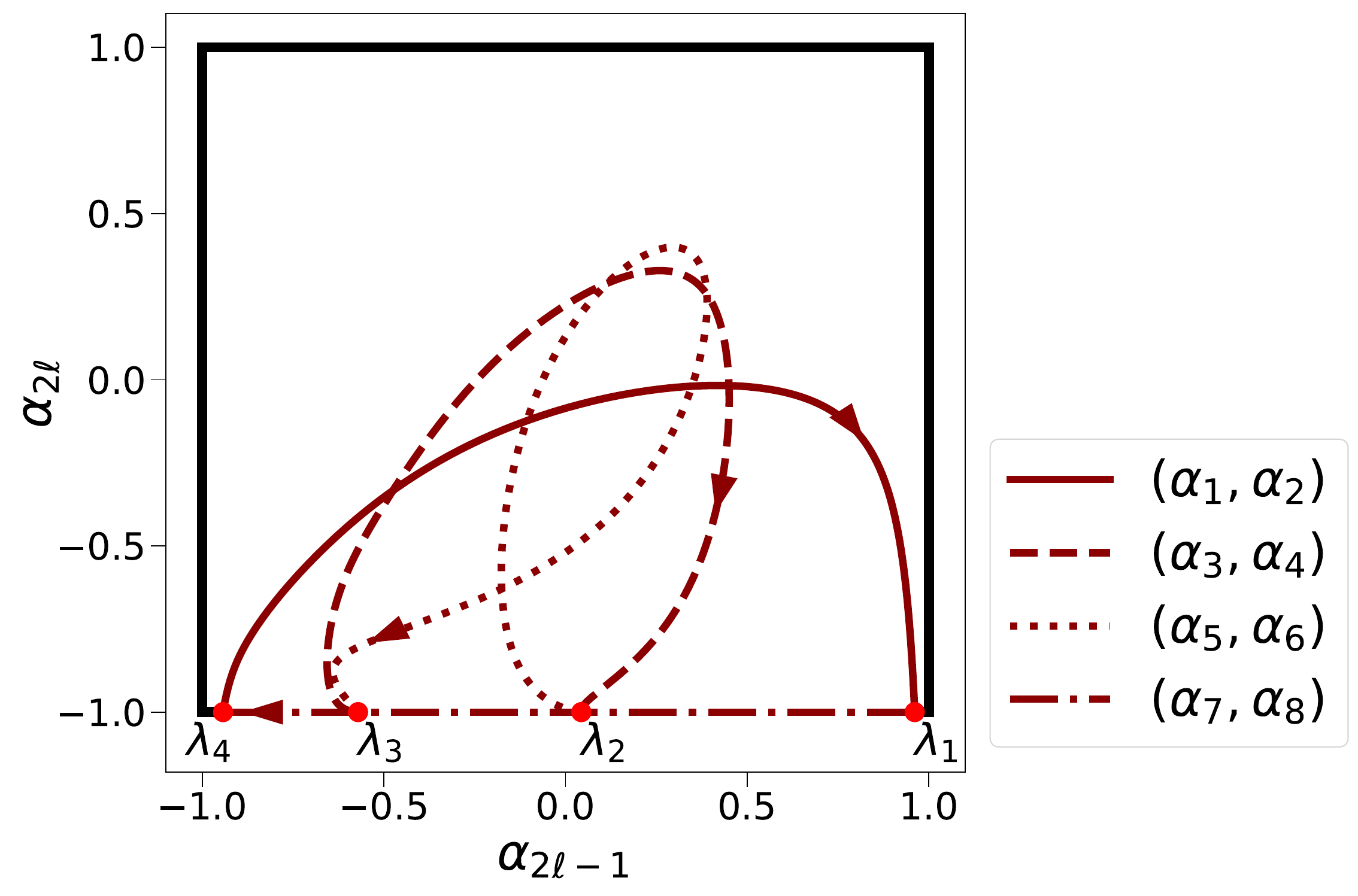}
    \caption{Numerical realization of a phase space plot for the Schur flow with $N=8$ and boundary conditions $\alpha_0=-1$, $\alpha_8=-1$. Since the $\alpha_j$ trajectories are on the interval $[-1,1]$, for graphical reasons, they are plotted in pairs, on the horizontal axis the ones with odd and on the vertical axis the ones with even label. Note that asymptotically $\alpha_{2l}$ goes to -1, both in the past and in the future. Special is the case of $\alpha_8(t)=-1$ for all $t$. Arrows point toward the future.} 
    \label{fig:MKDVphasespace}
\end{figure}
The even case is very similar to what happens for the Ablowitz-Ladik chain,
\beq\label{4.4}
\begin{split}
&\lim_{t\to+\infty}\alpha_{2\ell}(t)=(-1)
\qty
[1+\xi_{2\ell}\mathrm{e}^{-(\lambda_{\ell+1}-\lambda_{\ell})t}], \\
&\lim_{t\to-\infty}\alpha_{2\ell}(t)=(-1)
\qty
[1+\zeta_{2\ell}\mathrm{e}^{(\lambda_{M-\ell+1}-\lambda_{M-\ell})t}],
\end{split}
\eeq
where $\ell=1,\ldots,M$. The prefactors of the exponentials are given by
\beq\label{4.8}
\begin{split}
&\xi_{2\ell}=-\frac12\frac{(2\cos{\theta_{\ell+1}}-2\cos{\theta_{\ell}})^2}{(\sin{\theta_{\ell}})^2}\frac{\mu_{\ell+1}}{\mu_{\ell}}\prod_{k=1}^{\ell-1}
\abs
{\frac{2(\cos{\theta_{\ell+1}}-\cos{\theta_k})}{{2(\cos{\theta_{\ell}}-\cos{\theta_k})}}}^2, \\
&\zeta_{2\ell}=-\frac12\frac{(2\cos{\theta_{M-\ell+1}}-2\cos{\theta_{M-\ell}})^2}{(\sin{\theta_{M-\ell+1}})^2}\frac{\mu_{M-\ell}}{\mu_{M-\ell+1}}\prod_{j=M-\ell+2}^{M}
\abs
{\frac{2(\cos{\theta_{M-\ell}}-\cos{\theta_j})}{{2(\cos{\theta_{M-\ell+1}}-\cos{\theta_j})}}}^2.
\end{split}
\eeq
As before, the set $\{\mu_1,\ldots,\mu_N\}$ is the spectral measure of the Lax matrix $C$ with respect to the vector $(1,0,\dots,0)$.  However, due to the reflection-symmetry of the spectrum, the spectral measure is determined already by $\{\mu_1,\ldots,\mu_M\}$. 
As before, we define
 \begin{equation} \label{4.5}
q_{j+1}(t) - q_j(t) = r_j(t) = -\log \rho_j^2(t), \qquad \rho_j^2 = 1 - \alpha_j^2.
\end{equation}
Thus, the asymptotics \eqref{4.4} tells us $q_{2\ell+1}-q_{2\ell} = r_{2\ell} = -\log\rho_{2l}^2 \to +\infty$, hence, the two particles  are infinitely far apart at long times.
On the other hand, for the odd case the result is surprisingly different
\beq\label{4.6}
\begin{split}
&\lim_{t\to+\infty}\alpha_{2\ell-1}(t)=\tfrac{1}{2}\lambda_{\ell}, \\
&\lim_{t\to-\infty}\alpha_{2\ell-1}(t)=\tfrac{1}{2}\lambda_{M-\ell+1}.
\end{split}
\eeq
This has the striking consequence that  two-particle bound states are formed in the limit $|t| \to \infty$. In fact, $q_{2\ell}-q_{2\ell-1} = r_{2\ell-1} = -\log\rho_{2\ell-1}^2 \to -\log(1-(\lambda_{\ell}/2)^2)$, which implies that these two  particles maintain a constant distance at long times. The two particles labeled by the indices $(2\ell-1,2\ell)$  form a bound state moving with constant velocity. As will be seen, there is no need of extracting the exponential subleading corrections, contrary to the even case.
A numerical simulation of the trajectories inside the phase space is displayed in Figure \ref{fig:MKDVphasespace}.

Following the scheme used in the two previous sections, one wants to compute the analogue of the limit \eqref{3.15} to extract the scattering shift.
However, because of pair formation, one has to distinguish between odd and even labels.  
Thus the limit
\beq\label{4.10a}
\lim_{t\to+\infty} \rho_{2\ell-1}^2(t)\rho_{N-2\ell+1}^2(-t)= 
\qty
(1-\frac{\lambda_\ell^2}{4})^2
=(\sin^2{\theta_\ell})^2 = \exp{\Delta_{2\ell-1}-\Delta_{2\ell}}
\eeq
yields the exponential of the difference between the shifts of the two quasi-particles with labels $2\ell-1$ and $2\ell$ belonging to the same bound state. Notice that the term on the left carries no exponential factor, since the quasi-particles stay close to each other. The shift depends only on the rapidity of the two quasi-particle involved, 
\beq\label{4.10b}
\Delta_{2\ell-1}-\Delta_{2\ell} = 2\log\sin^2\theta_\ell,
\eeq
which amounts to the relative scattering shift $\log\sin^2\theta$. This however is not an effect due to scattering of the two quasi-particles inside the bound state. Rather it is caused by the reversed labelling at far past and future, compare with the comment below \eqref{2.9}. In particular, due to this reversal of labeling, a quasi-particle which was at the left of the bound state in the far past, would be at the right in the far future, acquiring a shift equal to the width of the bound state. According \eqref{4.6}, the length of a bound state of rapidity $\lambda_\ell$ is exactly $\log\sin^2\theta_\ell$.

On the other hand, considering  the limit
\beq\label{4.12}
\lim_{t\to+\infty} \rho_{2\ell}^2(t)\rho_{N-2\ell}^2(-t)\mathrm{e}^{2(\lambda_{\ell+1}-\lambda_\ell)t}=4\Re{\xi_{2\ell}}\Re{\zeta_{N-2\ell}} = \exp(\Delta_{2\ell}-\Delta_{2\ell+1})
\eeq
one obtains the difference between the scattering shift of the two quasi-particles $2\ell$ and $2\ell+1$ belonging to distinct bound states.
According to \eqref{4.8}, one finds 
\begin{eqnarray}
    \label{4.13}
&&\hspace{-30pt}4\Re{\xi_{2\ell}}\Re{N-\zeta_{2\ell}}\\
&&=\frac{(2\cos{\theta_{\ell+1}}-2\cos{\theta_{\ell}})^4}{(\sin^2{\theta_\ell})(\sin^2{\theta_{\ell+1}})}
\prod_{k=1}^{\ell-1}
\abs
{\frac{2(\cos{\theta_{\ell+1}}-\cos{\theta_k})}{{2(\cos{\theta_{\ell}}-\cos{\theta_k})}}}^2
\prod_{k=\ell+2}^{M}
\abs
{\frac{2(\cos{\theta_{\ell}}-\cos{\theta_k})}{{2(\cos{\theta_{\ell+1}}-\cos{\theta_k})}}}^2.\nonumber
\end{eqnarray}
Taking the log on both sides of this expression and recalling $\lambda_j=-2\cos\theta_j$, one obtains
\beq\label{4.14}
\begin{split}
&\Delta_{2\ell}-\Delta_{2\ell+1} = -\sum_{1 \leq k \leq M, k\neq \ell}\text{sign}(\lambda_\ell-\lambda_k)\log\abs{2
\qty(\cos\theta_\ell-\cos\theta_k)}^2 - \log\sin^2\theta_\ell \\
&+ \sum_{1 \leq k \leq M,k\neq \ell+1}\text{sign}(\lambda_{\ell+1}-\lambda_k)\log\abs{2
\qty(\cos\theta_{\ell+1}-\cos\theta_k)}^2  
 - \log\sin^2\theta_{\ell+1}.
\end{split}
\eeq
Thus, one ends up with
\beq\label{4.15}
\Delta_{2\ell} = -\sum_{1 \leq k \leq M,k\neq \ell}\text{sign}(\lambda_\ell-\lambda_k)\log\abs{2
\qty
(\cos\theta_\ell-\cos\theta_k)}^2 - \log\sin^2\theta_\ell.
\eeq
The quasi-particle with even label $2\ell$ and rapidity $\lambda_{\ell}=-2\cos\theta_\ell$ scatters with other bound pairs acquiring a shift equal to $\text{sign}(\lambda_k-\lambda_\ell)\log\abs{2
\qty
(\cos\theta_\ell-\cos\theta_k)}^2$. In addition, this quasi-particle acquires a shift $\log\sin^2\theta_\ell$ equal to the width of the bound state, due to the swap in position by reversing the labeling.
On the other hand, the quasi-particle with label $2\ell-1$ has shift
\beq\label{4.16}
\Delta_{2\ell-1} = -\sum_{1 \leq k \leq M,k\neq \ell}\text{sign}(\lambda_\ell-\lambda_k)\log\abs{2
\qty
(\cos\theta_\ell-\cos\theta_k)}^2 + \log\sin^2\theta_\ell.
\eeq
The scattering with the other bound state stays the same, while the internal shift has its sign reversed. 

Our discussion imposed the boundary conditions $\alpha_0=-1$ and $\alpha_N=-1$. Choosing instead  $\alpha_0=-1$ and $\alpha_N= 1$,  the rapidities become 
$\{-1,\lambda_2=\lambda_3<\cdots<\lambda_{N-2}=\lambda_{N-1},1\}$. Thus the even-odd indices are paired, leaving the first and and last index unpaired. As a modification of order 1, the TBA equations 
remain unaltered.
\subsection{Numerical solution, classical fusion, and TBA equation}
\label{sec4.2}
\begin{figure}
    \centering
    \includegraphics[width=\textwidth]{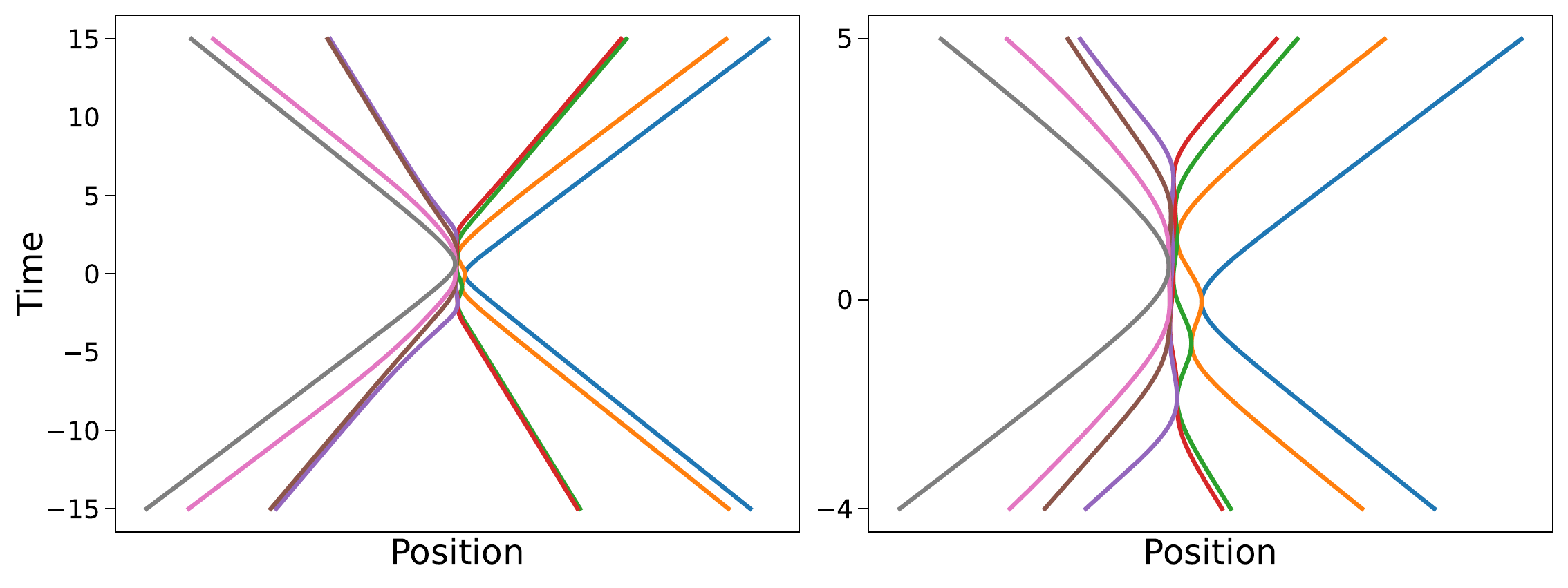}
    \caption{Example of particle trajectories for the Schur flow with $N=8$. On the left displayed is large time scale for which the quasi-particle picture appears neatly. Bound states can be clearly seen, especially for the faster particles. On the right represented is the short time scale with its intricate particle
    dynamics.}
    \label{fig:MKDVscattering}
\end{figure}
%
%
As in Section 4.2, we numerically solve the open chain \eqref{4.2} and infer the particle dynamics. Our result is plotted in Figure \ref{fig:MKDVscattering}. In the left panel bound states can be clearly identified. At least qualitatively, one also observes the width of the bound state to be proportional to the absolute value of its velocity, as derived in \eqref{4.6}.

From the many-body perspective, one key property of integrable systems is the factorization of scattering. As an important consequence, the scattering between bound states has to factorize as regards to the scattering between its constituents.
This is called \textit{fusion}\cite{T16}, lying at the heart of the string hypothesis introduced by 
Takahashi \cite{T71} in the context of the 1D Heisenberg spin chain in the thermodynamic limit.
The insights of Takahashi are paradigmatic for the solution of multi-component integrable models for which bound states between magnons appear as a rule\cite{BGI21,SCP21,SCP22}.  

The Schur flow is the first classical integrable many-particle model possessing such bound states. So let us check whether the fusion rule also holds in this case. 
In view of the results from Section \ref{sec3.1}, it is natural to conclude that 
the two-particle scattering shift for singletons is given by  \eqref{3.19}.
Let us then consider the scattering between a singleton and a bound pair. The three rapidities involved are $\mathrm{e}^{\mathrm{i}\theta},\mathrm{e}^{\mathrm{i}\theta'},\mathrm{e}^{-\mathrm{i}\theta'}$ and, according \eqref{4.15} and \eqref{4.16}, the singleton shift  is $\log|\mathrm{e}^{\mathrm{i}\theta}-\mathrm{e}^{\pm\mathrm{i}\theta'}|^2$. In fact, one observes that 
\beq
\label{4.17}
 \log|\mathrm{e}^{\mathrm{i}\theta}-\mathrm{e}^{\mathrm{i}\theta'}|^2 + \log|\mathrm{e}^{\mathrm{i}\theta}-\mathrm{e}^{-\mathrm{i}\theta'}|^2 = \log|2(\cos\theta-\cos\theta')|^2. 
\eeq
On the other hand, the shift due to the swap inside the bound pair, i.e. $\log|\sin\theta|^2$, cannot be factorized in the same way. In fact,
\beq\label{4.18}
 \log|2\sin\theta|^2 = \log|\mathrm{e}^{\mathrm{i}\theta}-\mathrm{e}^{-\mathrm{i}\theta}|^2 \neq \log|\sin\theta|^2,
\eeq
confirming that this term is not a scattering contribution.

For the AL chain we noted that the two-particle scattering shows up in the finite $N$ density of states under some GGE. A similar construction can be carried through for the Schur flow \cite{GM21,S21}.
The $M$ eigenvalues turn out to have the joint density
\beq 
\label{4.19}
\frac{1}{Z} |\Delta(2 \cos \theta_1,\ldots,2 \cos \theta_M)|^{2\mathsfit{P}/M}
 \prod_{j=1}^{M}(\sin\theta_j)^{(-1 + (P/M))}\mathrm{d}\theta_j.
\eeq 
The Vandermonde determinant $\Delta$ is the result of a computation.
But now its appearance can be understood as consequence of the fusion rule.

Relying on arguments similar to the ones in Section \ref{sec3.2} and using \eqref{4.19}, the generalized free energy has been computed.
The respective TBA equation then reads
\beq
\label{4.20}
\epsilon(\theta) = V(\theta) - \mu + \log\sin \theta  - 2\int_{0}^{\pi} \mathrm{d}\theta'\log(2|\cos \theta - \cos \theta'|)e^{-\varepsilon(\theta')}.
\eeq
As expected, the dressing operator is the two-particle scattering shift.

However, there is also the spurious addition $\log\sin \theta$ to the confining potential $V$ and one might guess a link to the one appearing in \eqref{4.15}. But the actual origin is more subtle. We start from \eqref{3.26}. In the limit $\mathsfit{P} \to 0$, the normalized a priori measure converges to the uniform distribution on the unit circle $\{z||z|=1\}$. Since $\rho_j \to 0$, the Lax matrix  
becomes diagonal, $C_{i,j} = -\delta_{ij}\alpha_{j-1}\bar{\alpha}_{j} $. Thereby one arrives at \eqref{3.28} for $\mathsfit{P} = 0$. The same reasoning can be applied to the Schur free energy leading to
the result that all eigenvalues are concentrated near $\pm 1$. But in actual fact, as stated in \eqref{4.19}, this limit turns out to be more singular and formally yields the a priori measure proportional to $|\sin\theta|^{-1} \mathrm{d}\theta$.

\section{Conclusions and outlook}
For classical many-particle systems the center of mass moves with constant velocity, which then permits the switch to relative/internal coordinates. 
We discovered a set of canonical coordinates for the complex discrete AL field $\alpha_j$ having the same flavor. The $r_j$'s are positional 
differences and the angles $\theta_j$ take the role of canonical momenta. 
The resulting Hamiltonian particle dynamics is fairly similar to the one of the Toda lattice. 
However, while Toda particles may cross each other, the AL particles maintain their order. Also for the AL system there is no separation in kinetic and potential energy, a property also familiar from classical spin systems, see e.g. \cite{DKH20}. In the AL two-particle problem,
the particles repel each other and switch to linear motion exponentially fast. For the AL particle system we establish rigorously that the relative scattering shifts for $N$ particles are determined by the two-particle scattering shift. Our results rely on the Lax matrix discovered by Nenciu \cite{N05} and the analysis of Killip and Nenciu  
\cite{KN07}. 

We also studied the Schur flow of the AL chain, which has the total momentum as generator of the dynamics. Completely unexpected, particles now form pairs asymptotically with a size depending on their velocity. For integrable quantum many-body systems the formation of 
bound states is familiar and well studied \cite{F17}, but for classical models such behavior has not been reported before. In quantum models bound states have a scattering shift governed by the fusion rule \cite{T71}. The same rule determines the scattering shift for the Schur flow.

As byproduct we confirm that the TBA dressing operator 
agrees with the two-particle scattering shift. While this is expected from the experience with integrable quantum many-body  systems, such result is remarkable in the classical context, since the identity relies on two entirely different methods. TBA is related to the generalized free energy and used are methods from random matrix theory in particular the CUE and COE ensemble, while scattering relies on a method developed by Moser in the context of the Toda lattice.

Only a few illustrative plots have been presented. But the AL particle dynamics has the potential of simulating large 
number of particles, thereby offering the possibility to check the predictions of GHD in the spirit of \cite{MS15,GMK23,BCdN23,SDSK23,M24}. Another direction would be to extend our results to other integrable many-particle systems. 
So far Toda, AL, and Calogero fluid are well covered.
A substantial list of further classical models is presented in \cite{GGG23} and first steps have been accomplished.
A very natural direction worth exploring is to link our study to the continuum nonlinear Schr\"odinger equation\cite{BdL20,KCB2022}. In particularly, it would be interesting to understand how to obtain the TBA equation for the defocusing NLS\cite{BdL20} as continuum limit of the AL chain. 

A major challenge poses the focusing AL chain, which in the equations of motion merely amounts to  a change from  $\rho_j^2 = 1 - \abs{\alpha_j}^2$ to $\rho_j^2 = 1 + \abs{\alpha_j}^2$. Then $\alpha_j \in \mathbb{C}$, which presumably changes the dynamics drastically. 
On the other hand, for algebraic purposes, one can regard $\alpha_j$ and $\bar{\alpha}_j$ as independent variables. The extra minus sign from before is then taken care off upon substituting $\alpha_j$ by $-\alpha_j$. Thus the algebraic identities leading to the Lax matrix \eqref{3.5} are still valid and, in particularly, also
the formulas for the conserved charges. However these charges are unbounded 
to both sides. Hence the usual grand-canonical prescription for GGE can no longer be applied.  
Only the a priori product measure is well-defined.

\section*{Acknowledgments}
The work is funded by the Deutsche Forschungsgemeinschaft (DFG, German Research Foundation) – TRR 352 – Project-ID 470903074. AB thanks the organizers of the workshop \textit{"Emergent Hydrodynamics of Integrable Systems and Soliton Gases"} at CIRM in Marseille, for the useful discussions of that week. We thank Alvise Bastianello and Christian B. Mendl for valuable comments and support.

\appendix
\setcounter{equation}{0}
\section{Two-particle dynamics and scattering shift}
For integrable particle systems, usually the scattering shift can be computed directly from the motion of two particles 
with open boundary conditions. As example, for the open Toda chain the two-particle motion is governed by 
\begin{equation} 
\label{0}
\ddot{q}_1 = -\mathrm{e}^{q_1-q_2},\quad \ddot{q}_2 = \mathrm{e}^{q_1-q_2},
\end{equation}
 see Sec. \ref{sec2}. Also for integrable quantum many-body systems one relies on the two-particle phase shift, as can be seen from the book of Sutherland who devotes a lengthy appendix in collecting solutions of two-body problems \cite{SU04}. With prior methods, for the AL lattice this route  seemed to be blocked. 
 However, having Section 2 at our disposal, such an approach can be followed and should result with the scattering shift \eqref{3.19}.
 Thus we consider $N=2$ and boundary phases
$\alpha_0 = 1, \alpha_2 = \mathrm{e}^{\mathrm{i}\vartheta}$ with the convention $-\pi\leq\vartheta \leq\pi$. For notational simplicity, the dynamical variable is 
$\alpha_1 = \alpha = \sqrt{(1-\mathrm{e}^{-r})}\mathrm{e}^{\mathrm{i}\varphi}$.
The Hamiltonian reads, see Sec. 2,
\begin{equation} 
\label{1}
H = - 2\sqrt{(1-\mathrm{e}^{-r})}(\cos\varphi+  \cos(\varphi -\vartheta))
\end{equation}
 with equations of motion
 \begin{eqnarray} 
\label{2}
&&\dot{r} =  2\sqrt{(1-\mathrm{e}^{-r})}(\sin\varphi+  \sin(\varphi -\vartheta)) = 
4\cos(\tfrac12\vartheta)\sqrt{1-e^{-r}}\sin(\varphi-\tfrac12\vartheta),\nonumber\\
&& \dot{\varphi} = \frac{\mathrm{e}^{-r}}{\sqrt{(1-\mathrm{e}^{-r}})} (\cos\varphi+  \cos(\varphi -\vartheta)).
\end{eqnarray}
 The particle positions are given through
 \begin{equation} 
\label{3}
\dot{q}_1 = - 2\sqrt{(1-\mathrm{e}^{-r})}\sin\varphi,  \quad \dot{q}_2 = 2 \sqrt{(1-\mathrm{e}^{-r})}\sin(\varphi - \vartheta).
\end{equation}
Since the energy is conserved, trajectories are constrained to $\{H = E\}$, to say
\begin{equation} 
\label{4}
E^2 = 4(1-\mathrm{e}^{-r})(2\cos(\tfrac{1}{2}\vartheta))^2 (1- \sin^2(\varphi - \tfrac{1}{2}\vartheta)).
\end{equation}
Inserting in \eqref{2} one obtains 
\begin{equation} 
\label{5}
\dot{r} = a\sqrt{ b- \mathrm{e}^{-r}},  
\end{equation}
with
\begin{equation} 
\label{6}
a = 4\cos(\tfrac{1}{2}\vartheta) \geq 0, \quad b = \big(1- a^{-2}E^2\big)^{1/2}.
\end{equation}

To fix the origin of time, we impose $\dot{r}(0) = 0$, implying that $b \geq \mathrm{e}^{-r(t)}$ for all $t$.  Setting $x = \mathrm{e}^{-r}$ results in
\begin{equation} 
\label{7}
\dot{x} = - a x\sqrt{b-x}. 
\end{equation}
The solution is
\begin{equation} 
\label{8}
x(t) = b \cosh^{-2}(\tfrac{1}{2}a \sqrt{b} t), 
\end{equation}
as can be confirmed directly. With this result one infers the long time asymptotics,
up to exponentially small terms, as
\begin{equation} 
\label{9}
r(t) = -\log (4b) + a\sqrt{b} t
\end{equation}
for $t \to +\infty$ and 
\begin{equation} 
\label{10}
r(t) = -\log (4b) - a\sqrt{b} t
\end{equation}
for $t \to -\infty$. 

We know already that for large $|t|$ the $\alpha$-trajectory asymptotically reaches the boundary of the unit disk $\alpha=\mathrm{e}^{\mathrm{i}\phi_\infty}$.
By conservation of energy, the limiting phase $\varphi_\infty$ is determined by
\begin{equation} 
\label{11}
E  = - 2(\cos\varphi_\infty +\cos(\varphi_\infty- \vartheta)). 
\end{equation}
This equation has two solutions denoted by 
\begin{equation} 
\label{12}
\varphi_+ = -\beta + \vartheta,\quad \varphi_- = \beta 
\end{equation}
for arbitrary $\beta \in [0,2\pi]$, thereby parametrizing the energy $E$.

From \eqref{3}
\begin{equation} 
\label{16}
\dot{q}_1(\infty) = - 2\sin\varphi_\pm, \quad \dot{q}_2(\infty) = 2  \sin(\varphi_\pm -\vartheta),
\end{equation}
Since $q_1 < q_2$, one has $\dot{q}_1(-\infty) >0$,
$\dot{q}_1(\infty) <0$, $\dot{q}_2(-\infty) <0$,  $\dot{q}_2(\infty) >0$, which fixes of how to choose 
$\varphi_+$ and $\varphi_-$. 

According to \eqref{9},  \eqref{10}, the relative two-particle scattering shift is $-\log(4b) $. Now
\begin{equation} 
\label{17}
\dot{r}(\infty) = a \sqrt{b}  = 2(\sin\varphi_\pm + \sin(\varphi_\pm -\vartheta)) = 4 \cos(\tfrac{1}{2}\vartheta)
\sin(\varphi_\pm -\tfrac{1}{2}\vartheta),
\end{equation}
and, because of  \eqref{6}, one gets $b=\sin^2(\varphi_\pm -\frac12\vartheta)$, implying the scattering shift
\begin{equation} 
\label{18}
- \log(4b) = -\log ( 4 \sin^2(\beta -\tfrac{1}{2}\vartheta)) = -\log|\mathrm{e}^{-\mathrm{i}\beta} - \mathrm{e}^{-\mathrm{i}(-\beta+\vartheta)}|^2,
\end{equation}
which indeed agrees with the AL scattering shift \eqref{3.19}. 

It is still instructive to determine the eigenvalues of the Lax matrix, since they have to match with the two terms in the argument of the log in the right hand side of \eqref{18}. One has
\begin{equation}\label{19} 
L =
\begin{pmatrix}
\bar{\alpha}& \rho \\
\rho & -\alpha \\
\end{pmatrix},\quad
M =
\begin{pmatrix}
-1& 0 \\
0 & \mathrm{e}^{-\mathrm{i}\vartheta} \\
\end{pmatrix},\quad
C = LM =
\begin{pmatrix}
-\bar{\alpha}& \rho\, \mathrm{e}^{-\mathrm{i}\vartheta} \\
-\rho& -\alpha\,\mathrm{e}^{-\mathrm{i}\vartheta}  \\
\end{pmatrix}.
\end{equation}
The eigenvalues of $C(t)$ are time-independent and can be easily determined from $C(\infty)$. Since  $\alpha(\infty)=\mathrm{e}^{\mathrm{i}\varphi_\pm}$ and $\rho(\infty)=0$, one obtains
\begin{equation}
    C(\infty)=
    \begin{pmatrix}
-e^{-i\varphi_\pm}& 0 \\
0& -e^{i(\varphi_\pm-\vartheta)}  \\
\end{pmatrix}=
    \begin{pmatrix}
-\mathrm{e}^{-\mathrm{i}\beta}& 0 \\
0& -\mathrm{e}^{-\mathrm{i}(-\beta+\vartheta)}  \\
\end{pmatrix}.
\end{equation}

\section{Asymptotics for the open AL hierarchy}\label{app:A}
In greater generality, one can study the dynamics of the AL model governed by the Hamiltonian $H_\mathsfit{f} = \Im\Tr(\mathsfit{f}(C))$ for some polynomial complex-valued function $\mathsfit{f}$. This is usually called the AL hierarchy. The long time behavior of the open hierarchy was studied by Killip and Nenciu \cite{KN07}. The purpose of this appendix is twofold. For convenience we state verbatim, including the numbering,  the for us relevant identities and theorems from \cite{KN07}.  While the results from Section \ref{sec3.1}
are a fairly direct consequence, the case of a reflection-symmetric spectrum 
requires additional arguments, which will be provided here.

For the Hamiltonian $H_\mathsfit{f}$, the rapidities are given by $\lambda_j=\Re[z_j\mathsfit{f}'(z_j)]$, where $\{z_1,\ldots,z_N\}$ are the eigenvalues of $C(t)$. Due to the Lax structure, the eigenvalues of $C(t)$ do not depend on time, while the spectral measure $\mu_j(t)$, as defined in \eqref{3.12}, does so. To be noted is the change of notation between this Appendix and Section 5. In the latter,  removing degeneracy, the $\lambda_j$'s run from $1$ to $M= N/2$,  which lightens our notation. Contrarily, in this appendix  $j=1,\ldots,N$ allowing $\lambda_j=\lambda_{j'}$ for some $j'\neq j$. 

For the following it will be useful to order the rapidities as $\lambda_N > \ldots > \lambda_1$ and introduce a multi-index notation.
Let us consider an ordered multi-index $I=(i_1<\cdots<i_k)$ of length $k$ and introduce the Vandermonde determinant,
\beq
\label{A.1}
\Delta_I=\Delta(z_{i_1},\ldots,z_{i_k})=\prod_{1\leq\ell<m\leq k} \big(z_{i_m}-z_{i_\ell}\big),
\eeq
and the products
\beq
\label{A.1a}
z_I=\prod_{i_\ell\in I} z_{i_\ell}, 
\qquad \mu_I(t)=\prod_{i_\ell\in I} \mu_{i_\ell}(t).
\eeq
In case $I$ consists of a single index, one sets $\Delta_I = 1$.
With this notation, the starting point of the analysis  is an exact expression, valid for any time,
    \beq\label{A.2}
    \alpha_j(t) = (-1)^j \frac{\sum \abs{\Delta_I}^2\mu_L(t)\bar{z}_I}
    {\sum \abs{\Delta_I}^2\mu_I(t)},
    \eeq
    where the sum is over all multi-indices $I$ of length $j$.

 To determine the long time asymptotics, one needs some information on the spectral measure, which is provided by\\\\   
PROPOSITION 7.1 \textit{
Under the flow generated by the Hamiltonian $H_\mathsfit{f}$, the spectral measures have the following asymptotics
\beq\label{A.3}
\log[\mu_j(t)] = -(\lambda_j-\lambda_1)t + \log\qty[\frac{\mu_j}{\mu_1+\ldots+\mu_\nu}] + \order{e^{-at}}
\eeq
as $t\to+\infty$. Here $\nu$ is defined by $\lambda_1=\ldots=\lambda_\nu < \lambda_{\nu+1}$, $a=\lambda_{\nu+1}-\lambda_1>0$ and time dependence is omitted when referring to initial data. In particular, if $j>\nu$, then $\mu_j(t)\to0$ exponentially fast.}\\

Compared to \cite{KN07}, for our results  time is reversed, i.e. our $t\to+\infty$ agrees with the results of Nenciu and Killip for $t\to-\infty$. The validity of our convention can be easily checked numerically. Most likely this can be traced to a different convention for the Poisson brackets.

With the input of Proposition 7.1, the main theorem reads\\\\
THEOREM 7.3 \textit{Fix $1\leq j \leq N$. Let $B_j=\{\ell: \lambda_\ell=\lambda_j\}$, and let $s(j)=\min B_j-1$. Then, as $t\to+\infty$,}
    \beq\label{A.4}
    \alpha_j(t)\to(-1)^{j-1}\bar{z}_J\frac{\sum\abs{\Delta_{J\cup I}}^2\mu_I\bar{z}_I}{\sum\abs{\Delta_{J\cup I}}^2\mu_I},
    \eeq
    \textit{where $J=(1<\cdots<s(j))$ and both sums are over all ordered multi-indices $I\subseteq B_j$ of length $j-s(j)$. In particular, $\abs{\alpha_j}\to1$ if and only if $\lambda_j>\lambda_{j-1}$.}

    \textit{If all $\lambda_j$ are distinct, then}
    \beq
    \label{A.5}
    \lim_{t \to +\infty} \alpha_j(t)=(-1)^{j-1} \Bar{z}_1\dots\Bar{z}_{j}
    \qty[1+\xi_{j}\mathrm{e}^{-(\lambda_{j+1}-\lambda_{j})t}+\order{\mathrm{e}^{-\gamma t}}],
    \eeq
    \textit{where}
    \beq
    \label{A.6}
    \xi_j=(z_{j}\Bar{z}_{j+1}-1)\frac{\mu_{j+1}}{\mu_{j}}\prod_{k=1}^{j-1}\abs{\frac{z_{j+1}-z_k}{z_{j}-z_k}}^2
    \eeq
    \textit{and $\gamma>(\lambda_{j+1}-\lambda_{j})$.}\\

To arrive at \eqref{A.5}, one has to identify the asymptotically dominant term in expressions as \eqref{A.4}.
    Exponentiating Eq. \eqref{A.3}, one gets
    \beq\label{A.7}
    \mu_L(t) = \frac{\mu_L}{(\mu_1+\ldots+\mu_{\nu})^j}\exp\Big(-t\sum( \lambda_{i_\ell}-\lambda_1)\Big)
    \eeq
    plus some exponential corrections which have been suppressed. Thus, the leading term is given by the multi-index which minimize $\lambda_{i_1} + \ldots + \lambda_{i_\ell}$. Since the labeling is in increasing order, the sum is minimal for $\lambda_1 + \ldots + \lambda_{j}$ and the set of multi-indices which achieve this value is exactly the collection of $J\cup I$ as stated in the theorem.
    This leads to
    \beq
    \label{A.8}
    \alpha_j(t)\to(-1)^{j-1}\frac{\sum\abs{\Delta_{J\cup I}}^2\mu_{J\cup I}(t)\bar{z}_{J\cup I}}{\sum\abs{\Delta_{J\cup I}}^2\mu_{J\cup I}(t)}.
    \eeq
    Since the sums are over $I$, one can factor out  $\bar{z}_J$ and simplify  $\mu_J(t)$, obtaining the result stated in the theorem.
    From \eqref{A.4} we infer that the limiting value of $(-1)^{j-1}z_J\alpha_j(t)$ is a convex combination of points on the unit circle. This sum consists of a single term if and only if $\lambda_j>\lambda_{j-1}$. Therefore  $\abs{\alpha_j(t)}\to 1$ under this condition.

    Let us now consider the case of distinct $\lambda_j$. We are looking for the first subleading term and want to find the two smallest values of $\lambda_{i_1} + \ldots + \lambda_{i_\ell}$. As before the first one is $\lambda_1 + \ldots + \lambda_{j-1} + \lambda_{j}$, while the second one $\lambda_1 + \ldots + \lambda_{j-1} + \lambda_{j+1}$. Inserting these two multi-indices in \eqref{A.4} obtains
    \beq
    \label{A.9}
    \begin{split}
    \alpha_j(t) \to &(-1)^{j-1} \Bar{z}_1\cdots\Bar{z}_{j}\frac{1+z_{j}\bar{z}_{j+1}P_je^{-(\lambda_{j+1}-\lambda_{j})t}+\order{e^{-\Tilde{\gamma}t}}}{{1+P_je^{-(\lambda_{j+1}-\lambda_{j})t}+\order{e^{-\Tilde{\gamma}t}}}}\\
    &= (-1)^{j-1} \Bar{z}_1\dots\Bar{z}_{j}\qty[1+(z_{j}\bar{z}_{j+1}-1)P_je^{-(\lambda_{j+1}-\lambda_{j})t} + \order{e^{-\gamma t}}],
    \end{split}
    \eeq
    where $\tilde{\gamma}=\lambda_j-\lambda_{j-2}$ and $\gamma=\min\{\Tilde{\gamma},2(\lambda_j-\lambda_{j-1})\}$, 
    and 
    \beq
    \label{A.10}
    P_j = \frac{\mu_{j+1}}{\mu_{j}}\abs{\frac{\Delta(z_{1},\ldots,z_{j-1},z_{j+1})}{\Delta(z_{1},\ldots,z_j)}}.
    \eeq

Theorem 7.3 asserts the leading asymptotic contribution for any system in the Ablowitz-Ladik hierarchy, with no assumptions on the degeneracy of the rapidities. Moreover, in the non-degenerate case it explicitly yields the exponential asymptotic decay to the limit value.
As explained in Section \ref{sec3.1}, with this input one arrives at the $N$-particle scattering shift. To deal with the Schur flow, we have to extend Theorem 7.3 to the case of a reflection-symmetric spectrum.\\\\
\textbf{Proposition}.\textit{
    Let the rapidities be two-fold degenerate with $\lambda_M > \ldots > \lambda_1$ and $2M=N$. Then}
    \beq\label{A.11}
    \begin{split}
    &\lim_{t\to+\infty}\alpha_{2\ell}(t)=(-1)
\qty
[1+\xi_{2\ell}\mathrm{e}^{-(\lambda_{\ell+1}-\lambda_{\ell})t}],\\
    &\lim_{t\to+\infty}\alpha_{2\ell-1}(t)=\tfrac{1}{2}\lambda_{\ell}, \\
    \end{split}
    \eeq
    \textit{where}
    \beq
    \label{A.12}
    \xi_{2\ell}=-\frac12\frac{(2\cos{\theta_{\ell+1}}-2\cos{\theta_{\ell}})^2}{(\sin{\theta_{\ell}})^2}\frac{\mu_{\ell+1}}{\mu_{\ell}}\prod_{k=1}^{\ell-1}
\abs
{\frac{2(\cos{\theta_{\ell+1}}-\cos{\theta_k})}{{2(\cos{\theta_{\ell}}-\cos{\theta_k})}}}^2.
    \eeq
\begin{proof} 
    We start from \eqref{A.2} and have to understand which multi-indices $I$ yield the leading and subleading contributions.
    In case of twofold degeneracy one has to distinguish whether the index is even or odd.
    
For even $j=2\ell$, the leading contribution is given by a unique multi-index set $I=(1 < \cdots < j)$ which implies that the rapidities from $\lambda_1$ to $\lambda_{\ell}$, with $2M=N$ and $j=2\ell$, have to be taken into account. This results in four subleading terms, given by substituting one of the two rapidities $\lambda_{\ell}$ by one of the two $\lambda_{\ell+1}$. A procedure similar to the one used in \eqref{A.5} yields the asymptotics of  $\alpha_{2\ell}(t)$ as stated in \eqref{A.11}.

On the contrary, if $j$ is odd, hence $j=2\ell-1$, there are two leading terms. The $2\ell-1$ largest rapidities are given by twice the rapidities from $\lambda_1$ to $\lambda_{\ell-1}$, and just once the rapidity $\lambda_{\ell}$. Hence, this choice is achieved in two distinct ways, since two indices correspond to the rapidity $\lambda_{\ell}$. The presence of two leading terms in the sum is the reason  for $|\alpha_{2\ell-1}(t)|$ not converging to $1$. Actually, in \eqref{A.2} we have two multi-indices $L$ providing the leading term, which are $I^{(1)} = (1 < \cdots < j+1)$ and $I^{(2)} = (1 < \cdots < j-1 < j+1)$. Noticing that $
\abs
{\Delta_{I^{(1)}}}^2$ = $
\abs
{\Delta_{I^{(2)}}}^2$ and that $\mu_{I^{(1)}}(t)$ and $\mu_{I^{(2)}}(t)$ have the same asymptotics, one obtains
\beq\label{A.13}
\begin{split}
\alpha_{2\ell-1}(t) &= (-1)^{2\ell-1}  \frac{
\abs
{\Delta_{I^{(1)}}}^2\mu_{I^{(1)}}(t)\bar{z}_{L^{(1)}} + 
\abs
{\Delta_{I^{(2)}}}^2\mu_{I^{(2)}}(t)\bar{z}_{I^{(2)}}}{
\abs
{\Delta_{I^{(1)}}}^2\mu_{I^{(1)}}(t) + 
\abs
{\Delta_{I^{(2)}}}^2\mu_{I^{(2)}}(t)}	\\
& = (-1)
\qty
(z_{\ell} + \bar{z}_{\ell}) = \frac{1}{2}\lambda_{\ell}
\end{split}
\eeq
for $t\to+\infty$.
\end{proof}

The subleading term in the asymptotics is required for the Moser-like limit when $\rho_j(t)^2\to 0$ as in the Ablowitz-Ladik case. In the Schur flow, we have $\rho_{2\ell-1}(t)^2$ different from zero at large time. Thus, in that case the Moser-like limit is meaningful also without subleading exponential corrections.

To obtain the limits for $t\to-\infty$, it suffices to perform the computations with $H_\mathsfit{f}$ replaced by $-H_\mathsfit{f}$ and correspondingly to $C$ replaced by $-C$.
A simple prescription to swap between the formulas for $t\to+\infty$ and the ones for $t\to-\infty$ is given by
\beq
\label{A.14}
\lambda_j\mapsto-\lambda_{N-j+1},    \qquad  z_j\mapsto z_{N-j+1}, \qquad
\mu_j\mapsto\mu_{N-j+1}.
\eeq


\bibliographystyle{JHEP}
\bibliography{refs.bib}

\providecommand{\href}[2]{#2}\begingroup\raggedright\begin{thebibliography}{10}

\bibitem{PSSV11}
A.~Polkovnikov, K.~Sengupta, A.~Silva and M.~Vengalattore, \emph{Colloquium:
  Nonequilibrium dynamics of closed interacting quantum systems},
  \href{https://doi.org/10.1103/revmodphys.83.863}{\emph{Reviews of Modern
  Physics} {\bfseries 83} (2011) 863–883}.

\bibitem{CDY16}
O.A.~Castro-Alvaredo, B.~Doyon and T.~Yoshimura, \emph{Emergent hydrodynamics
  in integrable quantum systems out of equilibrium},
  \href{https://doi.org/10.1103/PhysRevX.6.041065}{\emph{Phys. Rev. X}
  {\bfseries 6} (2016) 041065}.

\bibitem{BCDF16}
B.~Bertini, M.~Collura, J.~De~Nardis and M.~Fagotti, \emph{Transport in
  out-of-equilibrium $xxz$ chains: Exact profiles of charges and currents},
  \href{https://doi.org/10.1103/PhysRevLett.117.207201}{\emph{Phys. Rev. Lett.}
  {\bfseries 117} (2016) 207201}.

\bibitem{D19a}
B.~Doyon, \emph{Lecture notes on generalised hydrodynamics},
  \href{https://doi.org/10.21468/scipostphyslectnotes.18}{\emph{SciPost Physics
  Lecture Notes} (2020) }.

\bibitem{S23}
H.~Spohn, \emph{Hydrodynamic Scales of Integrable Many-Body Systems}, World
  Scientific (2024).

\bibitem{GHDbose}
I.~Bouchoule and J.~Dubail, \emph{Generalized hydrodynamics in the
  one-dimensional bose gas: theory and experiments}, {\emph{Journal of
  Statistical Mechanics: Theory and Experiment} {\bfseries 2022} (2021) }.

\bibitem{DGMSV23}
B.~Doyon, S.~Gopalakrishnan, F.~Møller, J.~Schmiedmayer and R.~Vasseur,
  \emph{Generalized hydrodynamics: a perspective}, {\emph{arXiv} (2023) }
  [\href{https://arxiv.org/abs/2311.03438}{{\ttfamily 2311.03438}}].

\bibitem{JSTAT24}
A.~Bastianello, B.~Bertini, B.~Doyon and R.~Vasseur, \emph{Introduction to the
  special issue on emergent hydrodynamics in integrable many-body systems},
  \href{https://doi.org/10.1088/1742-5468/ac3e6a}{\emph{Journal of Statistical
  Mechanics: Theory and Experiment} {\bfseries 2022} (2022) 014001}.

\bibitem{JPhysA24}
A.~Abanov, B.~Doyon, J.~Dubail, A.~Kamenev and H.~Spohn, \emph{Hydrodynamics of
  low-dimensional quantum systems},
  \href{https://doi.org/10.1088/1751-8121/acecc8}{\emph{Journal of Physics A:
  Mathematical and Theoretical} {\bfseries 56} (2023) 370201}.

\bibitem{EK05}
G.A.~El and A.M.~Kamchatnov, \emph{Kinetic equation for a dense soliton gas},
  \href{https://doi.org/10.1103/PhysRevLett.95.204101}{\emph{Phys. Rev. Lett.}
  {\bfseries 95} (2005) 204101}.

\bibitem{E21}
G.A.~El, \emph{Soliton gas in integrable dispersive hydrodynamics},
  \href{https://doi.org/10.1088/1742-5468/ac0f6d}{\emph{Journal of Statistical
  Mechanics: Theory and Experiment} {\bfseries 2021} (2021) 114001}.

\bibitem{BDE22}
T.~Bonnemain, B.~Doyon and G.~El, \emph{Generalized hydrodynamics of the kdv
  soliton gas}, \href{https://doi.org/10.1088/1751-8121/ac8253}{\emph{Journal
  of Physics A: Mathematical and Theoretical} {\bfseries 55} (2022) 374004}.

\bibitem{M75}
J.~Moser, \emph{Finitely many mass points on the line under the influence of an
  exponential potential -- an integrable system},  in \emph{Dynamical Systems,
  Theory and Applications: Battelle Seattle 1974 Rencontres}, J.~Moser, ed.,
  (Berlin, Heidelberg), pp.~467--497, Springer Berlin Heidelberg (1975).

\bibitem{T75}
S.~Tanaka, \emph{Korteweg-de vries equation; asymptotic behavior of solutions},
  {\emph{Publications of The Research Institute for Mathematical Sciences}
  {\bfseries 10} (1974) 367}.

\bibitem{T76}
E.~Date and S.~Tanaka, \emph{Periodic multi-soliton solutions of korteweg-de
  vries equation and toda lattice}, {\emph{Progress of Theoretical Physics
  Supplement} {\bfseries 59} (1976) 107}.

\bibitem{D08}
B.~Doyon, \emph{Introduction to integrable quantum field theory}, Lecture
  notes, Durham University (2008).

\bibitem{F17}
F.~Franchini, \emph{An Introduction to Integrable Techniques for
  One-Dimensional Quantum Systems}, Springer International Publishing (2017).

\bibitem{D19}
B.~Doyon, \emph{Generalized hydrodynamics of the classical toda system},
  \href{https://doi.org/10.1063/1.5096892}{\emph{Journal of Mathematical
  Physics} {\bfseries 60} (2019) }.

\bibitem{S19}
H.~Spohn, \emph{Generalized gibbs ensembles of the classical toda chain},
  \href{https://doi.org/10.1007/s10955-019-02320-5}{\emph{Journal of
  Statistical Physics} {\bfseries 180} (2019) 4–22}.

\bibitem{AL75}
M.J.~Ablowitz and J.F.~Ladik, \emph{Nonlinear differential difference
  equations}, \href{https://doi.org/10.1063/1.522558}{\emph{Journal of
  Mathematical Physics} {\bfseries 16} (1975) 598}.

\bibitem{AL76}
M.J.~Ablowitz and J.F.~Ladik, \emph{{Nonlinear differential difference
  equations and Fourier analysis}},
  \href{https://doi.org/10.1063/1.523009}{\emph{Journal of Mathematical
  Physics} {\bfseries 17} (1976) 1011}.

\bibitem{APT04}
M.J.~Ablowitz, B.~Prinari and A.D.~Trubatch, \emph{Discrete and Continuous
  Nonlinear Schrödinger Systems}, London Mathematical Society Lecture Note
  Series, Cambridge University Press (2003).

\bibitem{GM21}
T.~Grava and G.~Mazzuca, \emph{Generalized gibbs ensemble of the
  ablowitz–ladik lattice, circular $\beta $-ensemble and double confluent
  heun equation},
  \href{https://doi.org/10.1007/s00220-023-04642-8}{\emph{Communications in
  Mathematical Physics} {\bfseries 399} (2023) 1689–1729}.

\bibitem{S21}
H.~Spohn, \emph{Hydrodynamic equations for the ablowitz–ladik discretization
  of the nonlinear schrödinger equation},
  \href{https://doi.org/10.1063/5.0075670}{\emph{Journal of Mathematical
  Physics} {\bfseries 63} (2022) }.

\bibitem{KN07}
R.~Killip and I.~Nenciu, \emph{Cmv: The unitary analogue of jacobi matrices},
  \href{https://doi.org/https://doi.org/10.1002/cpa.20160}{\emph{Communications
  on Pure and Applied Mathematics} {\bfseries 60} (2007) 1148}.

\bibitem{T16}
S.J.v.~Tongeren, \emph{Introduction to the thermodynamic bethe ansatz},
  \href{https://doi.org/10.1088/1751-8113/49/32/323005}{\emph{Journal of
  Physics A: Mathematical and Theoretical} {\bfseries 49} (2016) 323005}.

\bibitem{H89}
A.~Hubacher, \emph{{Classical scattering theory in one dimension}},
  {\emph{Communications in Mathematical Physics} {\bfseries 123} (1989) 353 }.

\bibitem{K67}
C.S.~Gardner, J.M.~Greene, M.D.~Kruskal and R.M.~Miura, \emph{Method for
  solving the korteweg-devries equation},
  \href{https://doi.org/10.1103/PhysRevLett.19.1095}{\emph{Phys. Rev. Lett.}
  {\bfseries 19} (1967) 1095}.

\bibitem{N05}
I.~Nenciu, \emph{Lax pairs for the ablowitz-ladik system via orthogonal
  polynomialson the unit circle},
  \href{https://doi.org/10.1155/IMRN.2005.647}{\emph{International Mathematics
  Research Notices} {\bfseries 2005} (2005) 647}.

\bibitem{F10}
P.J.~Forrester, \emph{Log-Gases and Random Matrices}, Princeton University
  Press, Princeton (2010).

\bibitem{MM23}
G.~Mazzuca and R.~Memin, \emph{{Large deviations for Ablowitz-Ladik lattice,
  and the Schur flow}},
  \href{https://doi.org/10.1214/23-EJP941}{\emph{Electronic Journal of
  Probability} {\bfseries 28} (2023) 1 }.

\bibitem{G06}
L.B.~Golinskii, \emph{Schur flows and orthogonal polynomials on the unit
  circle},
  \href{https://doi.org/10.1070/sm2006v197n08abeh003792}{\emph{Sbornik:
  Mathematics} {\bfseries 197} (2006) 1145–1165}.

\bibitem{T71}
M.~Takahashi, \emph{{One-Dimensional Heisenberg Model at Finite Temperature}},
  \href{https://doi.org/10.1143/PTP.46.401}{\emph{Progress of Theoretical
  Physics} {\bfseries 46} (1971) 401}.

\bibitem{BGI21}
V.B.~Bulchandani, S.~Gopalakrishnan and E.~Ilievski, \emph{Superdiffusion in
  spin chains}, \href{https://doi.org/10.1088/1742-5468/ac12c7}{\emph{Journal
  of Statistical Mechanics: Theory and Experiment} {\bfseries 2021} (2021)
  084001}.

\bibitem{SCP21}
S.~Scopa, P.~Calabrese and L.~Piroli, \emph{Real-time spin-charge separation in
  one-dimensional fermi gases from generalized hydrodynamics},
  \href{https://doi.org/10.1103/PhysRevB.104.115423}{\emph{Phys. Rev. B}
  {\bfseries 104} (2021) 115423}.

\bibitem{SCP22}
S.~Scopa, P.~Calabrese and L.~Piroli, \emph{Generalized hydrodynamics of the
  repulsive spin-$\frac{1}{2}$ fermi gas},
  \href{https://doi.org/10.1103/PhysRevB.106.134314}{\emph{Phys. Rev. B}
  {\bfseries 106} (2022) 134314}.

\bibitem{DKH20}
A.~Das, K.~Damle, A.~Dhar, D.A.~Huse, M.~Kulkarni, C.B.~Mendl et~al.,
  \emph{Nonlinear fluctuating hydrodynamics for the classical xxz spin chain},
  \href{https://doi.org/10.1007/s10955-019-02397-y}{\emph{Journal of
  Statistical Physics} {\bfseries 180} (2020) 238}.

\bibitem{MS15}
C.B.~Mendl and H.~Spohn, \emph{Low temperature dynamics of the one-dimensional
  discrete nonlinear schrödinger equation},
  \href{https://doi.org/10.1088/1742-5468/2015/08/p08028}{\emph{Journal of
  Statistical Mechanics: Theory and Experiment} {\bfseries 2015} (2015) }.

\bibitem{GMK23}
G.~Mazzuca, T.~Grava, T.~Kriecherbauer, K.T.-R.~McLaughlin, C.B.~Mendl and
  H.~Spohn, \emph{Equilibrium spacetime correlations of the toda lattice on the
  hydrodynamic scale},
  \href{https://doi.org/10.1007/s10955-023-03155-x}{\emph{Journal of
  Statistical Physics} {\bfseries 190} (2023) 149}.

\bibitem{BCdN23}
L.~Biagetti, G.~Cecile and J.D.~Nardis, \emph{Three-stage thermalisation of a
  quasi-integrable system}, {\emph{arXiv} (2023) }
  [\href{https://arxiv.org/abs/2307.05379}{{\ttfamily 2307.05379}}].

\bibitem{SDSK23}
S.K.~Singh, A.~Dhar, H.~Spohn and A.~Kundu, \emph{Thermalization and
  hydrodynamics in an interacting integrable system: the case of hard rods},
  {\emph{arXiv} (2023) } [\href{https://arxiv.org/abs/2310.18684}{{\ttfamily
  2310.18684}}].

\bibitem{GGG23}
T.~Grava, M.~Gisonni, G.~Gubbiotti and G.~Mazzuca, \emph{Discrete integrable
  systems and random lax matrices},
  \href{https://doi.org/10.1007/s10955-022-03024-z}{\emph{Journal of
  Statistical Physics} {\bfseries 190} (2022) 10}.

\bibitem{BdL20}
G.D.V.D.~Vecchio, A.~Bastianello, A.D.~Luca and G.~Mussardo, \emph{{Exact
  out-of-equilibrium steady states in the semiclassical limit of the
  interacting Bose gas}},
  \href{https://doi.org/10.21468/SciPostPhys.9.1.002}{\emph{SciPost Phys.}
  {\bfseries 9} (2020) 002}.

\bibitem{KCB2022}
R.~Koch, J.-S.~Caux and A.~Bastianello, \emph{Generalized hydrodynamics of the
  attractive non-linear schr\"odinger equation},
  \href{https://doi.org/10.1088/1751-8121/ac53c3}{\emph{Journal of Physics A:
  Mathematical and Theoretical} {\bfseries 55} (2022) 134001}.

\end{thebibliography}\endgroup
\end{document}